\documentclass[fleqn,usenatbib]{mnras}

\usepackage{newtxtext,newtxmath}

\usepackage[T1]{fontenc}
\usepackage{ae,aecompl}

\usepackage{graphicx}	

\usepackage{amsmath}	
\usepackage{bm} 
\usepackage{ulem}
\usepackage{balance}
\usepackage{enumitem}
\usepackage[usenames,dvipsnames]{xcolor}

\usepackage{verbatim}

\def\be{\begin{equation}}
\def\ee{\end{equation}}
\def\bea{\begin{eqnarray}}
\def\eea{\end{eqnarray}}

\usepackage{geometry}
\hypersetup{
    colorlinks = true,
    citecolor = {MidnightBlue},
    linkcolor = {BrickRed},
    urlcolor = {BrickRed}
}

\author[S.~Choudhuri et al.]{
Samir Choudhuri,$^{1}$\thanks{E-mail: s.choudhuri@qmul.ac.uk}
Philip Bull$^{1,2}$
and
Hugh Garsden$^{1}$
\\
$^{1}$Astronomy Unit, Queen Mary University of London, Mile End Road, London E1 4NS, United Kingdom \\
$^{2}$Department of Physics and Astronomy, University of Western Cape, Cape Town 7535, South Africa
}

\date{Accepted XXX. Received YYY; in original form ZZZ}

\pubyear{2020}

\begin{document}
\title[Primary beam non-redundancy in 21cm observations]{Patterns of primary beam non-redundancy in close-packed \\
21cm array observations}
\maketitle

\begin{abstract}
Radio interferometer arrays such as HERA consist of many close-packed dishes arranged in a regular pattern, giving rise to a large number of `redundant' baselines with the same length and orientation. Since identical baselines should see an identical sky signal, this provides a way of finding a relative gain/bandpass calibration without needing an explicit sky model. In reality, there are many reasons why baselines will not be exactly identical, giving rise to a host of effects that spoil the redundancy of the array and induce spurious structure in the calibration solutions if not accounted for. In this paper, we seek to build an understanding of how differences in the primary beam response between antennas affect redundantly-calibrated interferometric visibilities and their resulting frequency (delay-space) power spectra. We use simulations to study several generic types of primary beam variation, including differences in the width of the main lobe, the angular and frequency structure of the sidelobes, and the beam ellipticity and orientation. 
For all of these types, we find that additional temporal structure is induced in the gain solutions, particularly when bright point sources pass through the beam. In comparison, only a low level of additional spectral structure is induced. The temporal structure modulates the cosmological 21cm power spectrum, but only at the level of a few percent in our simulations. We also investigate the possibility of signal loss due to decoherence effects when non-redundant visibilities are averaged together, finding that the decoherence is worst when bright point sources pass through the beam, and that its magnitude varies significantly between baseline groups and types of primary beam variation. Redundant calibration absorbs some of the decoherence effect however, reducing its impact compared to if the visibilities were perfectly calibrated.

\end{abstract} 


\section{Introduction}
\label{intro}

Detection of the 21cm line from neutral hydrogen promises to probe the dynamics, evolution, and thermal state of the Universe from the Dark Ages through to the present dark energy-dominated epoch. At some point in the intervening period, the gas content of the Universe changed phase from being completely neutral to almost fully ionised, through a process called reionisation. Many unresolved questions about this process, such as its exact timing and duration, and which astrophysical sources are responsible for it, can be answered by studying the evolution and clustering properties of the 21cm emission at redshifts between roughly $6 \lesssim z \lesssim 20$ \citep{furlanetto06,morales10,pritchard12,mellema13}.

Several ongoing and future experiments are primarily aimed at detecting the 21cm clustering signal from the Epoch of Reionisation (EoR), including the Giant Meterwave Radio Telescope (GMRT\footnote{\url{http://www.gmrt.ncra.tifr.res.in/}}; \citealt{swarup91,paciga11}), the Low Frequency Array (LOFAR\footnote{\url{http://www.lofar.org/}}; \citealt{vanhaarlem13,gehlot19,mertens20}), the Murchison Wide-field Array (MWA\footnote{\url{http://www.mwatelescope.org}}; \citealt{tingay13,trott20}), the Donald C. Backer Precision Array to Probe the Epoch of Reionization (PAPER\footnote{\url{http://eor.berkeley.edu/}};  \citealt{parsons10,kolopanis19}), the Hydrogen Epoch of Reionization Array (HERA\footnote{\url{http://reionization.org/}}; \citealt{DeBoer:2016tnn}) and the Square Kilometre Array (SKA;
\citealt{koopmans15}). 

Unfortunately, astrophysical foregrounds that are around 4--5 orders of magnitude brighter than the cosmological 21cm signal present a severe challenge for its detection \citep{santos05,ali08,ghosh12,choudhuri20}. To main approaches have been used to try and overcome this issue: (a) {\it Foreground removal}, which subtracts model foreground components and uses the resulting residual data for 21cm estimation \citep{jelic08,chapman12,hothi20}; and (b) {\it Foreground avoidance}, which discards a wedge-shaped region in the $(k_{\perp}, \, k_{\parallel})$ plane that the foregrounds should be localised within \citep{datta10,vedantham12,thyag13,liu14a,liu14b}. In order to accurately model and subtract these foregrounds, the instrument must first be calibrated precisely. However, due to a wide variety of instrumental, atmospheric, and modelling effects, the recovered instrumental calibration always deviates from its true values in real observations \citep[e.g.][]{datta09,datta10, Barry:2016cpg,2016MNRAS.463.4317P, 2017MNRAS.470.1849E, 2018MNRAS.478.1484G, 2019A&A...622A...5D, 2019ApJ...882...58K, 2019MNRAS.483.5480M, 2020arXiv200308399D,2020ApJ...890..122K, kumar20}.

In the traditional sky-based calibration approach, where standard calibrator sources are used to solve for the antenna gains, inaccuracies in the sky model contaminate the modes outside the foreground wedge in the power spectrum measurement, potentially causing a bias in the recovered EoR signal \citep{Barry:2016cpg, 2017MNRAS.470.1849E, kumar20}. It is therefore expected that a sky model-based calibration precision on the order of $\sim10^{-5}$ is needed to conclusively detect the 21cm signal \citep{Barry:2016cpg}, although baseline weighting schemes can help mitigate this requirement \citep{2017MNRAS.470.1849E}.

Alternatively, the radio telescope itself can be designed to facilitate accurate calibration. This is an important motivation for adopting a highly redundant array design, e.g. for the PAPER and HERA experiments, as in principle redundancy allows for more accurate relative calibration of antenna gains \citep{2010MNRAS.408.1029L, 2016ApJ...826..181D}. By measuring effectively the same mode on the sky multiple times, with many baselines, an over-constrained system of simultaneous equations can be written down that allows one to solve for the gain parameters and true visibilities without needing a priori knowledge of the sky brightness distribution. We can write the observed visibility for a baseline between antennas $i$ and $j$ as
\be \label{eq:gains}
V_{ij} = g_i g_j^* V_{ij}^{\,{\rm true}} + n_{ij},
\ee
where $g_i(\nu, t)$ are the complex gains for antenna $i$ at frequency $\nu$ and time $t$, $V_{ij}^{\,{\rm true}}$ is the visibility that would be observed with a perfectly calibrated instrument, and $n_{ij}$ is the noise {(note that we have taken the noise to be independent of the antenna gains in this expression).} Importantly, for a perfectly redundant array, all baselines with the same length and orientation will share the same true visibility, differing only by their complex gains and noise. For arrays with a high degree of redundancy, this greatly reduces the number of degrees of freedom that must be solved for during calibration, leading to corresponding improvements in estimates of the gains and true visibilities.

While compelling, the redundant calibration approach cannot be performed entirely without reference to a sky model \citep{2018ApJ...863..170L, 2020arXiv200308399D, 2020ApJ...890..122K}. Several degenerate degrees of freedom occur within the system of simultaneous equations for the gains and visibilities that cannot be solved for using redundant calibration alone, and must therefore be fixed by reference to an absolute calibration that fixes the overall flux scale, a phase reference, and several other degrees of freedom related to the orientation of the array \citep{2020arXiv200308399D}. {Imperfections in the absolute calibration, e.g. due to incompleteness of the sky model, can overwhelm the EoR 21cm signal \citep{byrne19}, making this step an important source of calibration systematics. We will not study absolute calibration systematics further in this paper however.}

More importantly for the purposes of this paper, practical construction and deployment of radio antennae always results in imperfections in the array, resulting in small deviations away from perfect redundancy. For example, feeds can be mis-aligned, rotated, or displaced from their ideal positions \citep{joseph18, Orosz:2018avj}, while primary beam patterns of antennae differ due to slight electronic or mechanical variations \citep{10.1093/mnras/sty2433}. Even if these imperfections can be kept within reasonably stringent tolerances, environmental effects such as ambient temperature changes and wind loading can affect each array element differently, giving rise to additional variations across the array. {Close-packed arrays are also subject to antenna position-dependent effects such as mutual coupling (or cross-talk) between neighbouring antennas, which also breaks perfect redundancy since antennas located at the edge of the array behave differently to antennas at the centre \citep[e.g. see][]{Fagnoni2021}.}

The ultimate impact of non-redundancies depends on the properties of the calibration method that is applied to the data. A method that assumes perfect redundancy will necessarily absorb some of the baseline-to-baseline variations caused by non-redundancy into the gain and calibrated visibility solutions. Depending on the source of the non-redundancy, {this can cause} spurious additional spectral structure that will interact with bright foregrounds. As shown in \cite{Orosz:2018avj}, this expands the size of the foreground wedge region, {particularly  at longer baseline lengths, and therefore reduces} the number of modes available for 21cm signal detection. {In a related effect, variations in antenna position can also cause a bias in the phase of the antenna gain solutions \citep{joseph18}.}

As part of a delay spectrum analysis \citep{Parsons:2012qh}, redundant visibilities can be coherently averaged before the power spectrum estimation step \citep[e.g.][]{2015ApJ...809...61A, Ali:2018erratum}. In a perfectly redundant setting, this would result in significant improvements in signal to noise as the noise will average down as the number of baselines, $N_{\rm bl}$, rather than the $\sqrt{N_{\rm bl}}$ scaling achieved by incoherently averaging the power spectra themselves. Gain errors caused by non-redundancy can lead to decoherence (partial cancellation) as the visibilities are averaged however, resulting in a loss of signal power and jeopardising the interpretation of upper limits on the EoR power spectrum \citep{2017MNRAS.470.1849E, kumar20}. Non-redundancies therefore pose a potentially serious threat to attempts to constrain the 21cm power spectrum from EoR and Cosmic Dawn with redundant arrays such as HERA \citep{DeBoer:2016tnn}, and potentially other experiments that use redundant calibration \cite[e.g. HIRAX,][]{2016SPIE.9906E..5XN}.

In this paper, we build on the analysis of \cite{Orosz:2018avj} to characterise the effects of various types of primary beam non-redundancy on a hexagonal close-packaged array with similar properties to HERA. As well as studying variations in the width of the main lobe of the primary beam, we include a more realistic model of the primary beam sidelobes and their possible variation across the array; variations in their ellipticity and orientation; and different distributions of the deviations from non-redundancy (e.g. purely random vs. distributions with outliers). We apply the redundant calibration method described in \cite{2020arXiv200308399D}, without any reference to a sky model for absolute calibration, and compare coherent- and incoherently-averaged power spectrum estimates for each type of non-redundancy.

The paper is organised as follows. In Section~\ref{sec:models} we describe our fiducial model for the array layout and primary beam, and set out the different types of deviation from perfect redundancy that we consider. In Section~\ref{sec:sims} we give an overview of our visibility simulations, and Section~\ref{sec:analysis} describes the synthetic calibration and  power spectrum estimation pipeline applied to them. In Section~\ref{sec:results} we analyse the effects of the different types of non-redundancy on the gain solutions and the coherent- and incoherently-averaged power spectra. Finally, we summarize and conclude in Section~\ref{discuss}.

\begin{figure}
\begin{center}
\includegraphics[width=1.01\columnwidth]{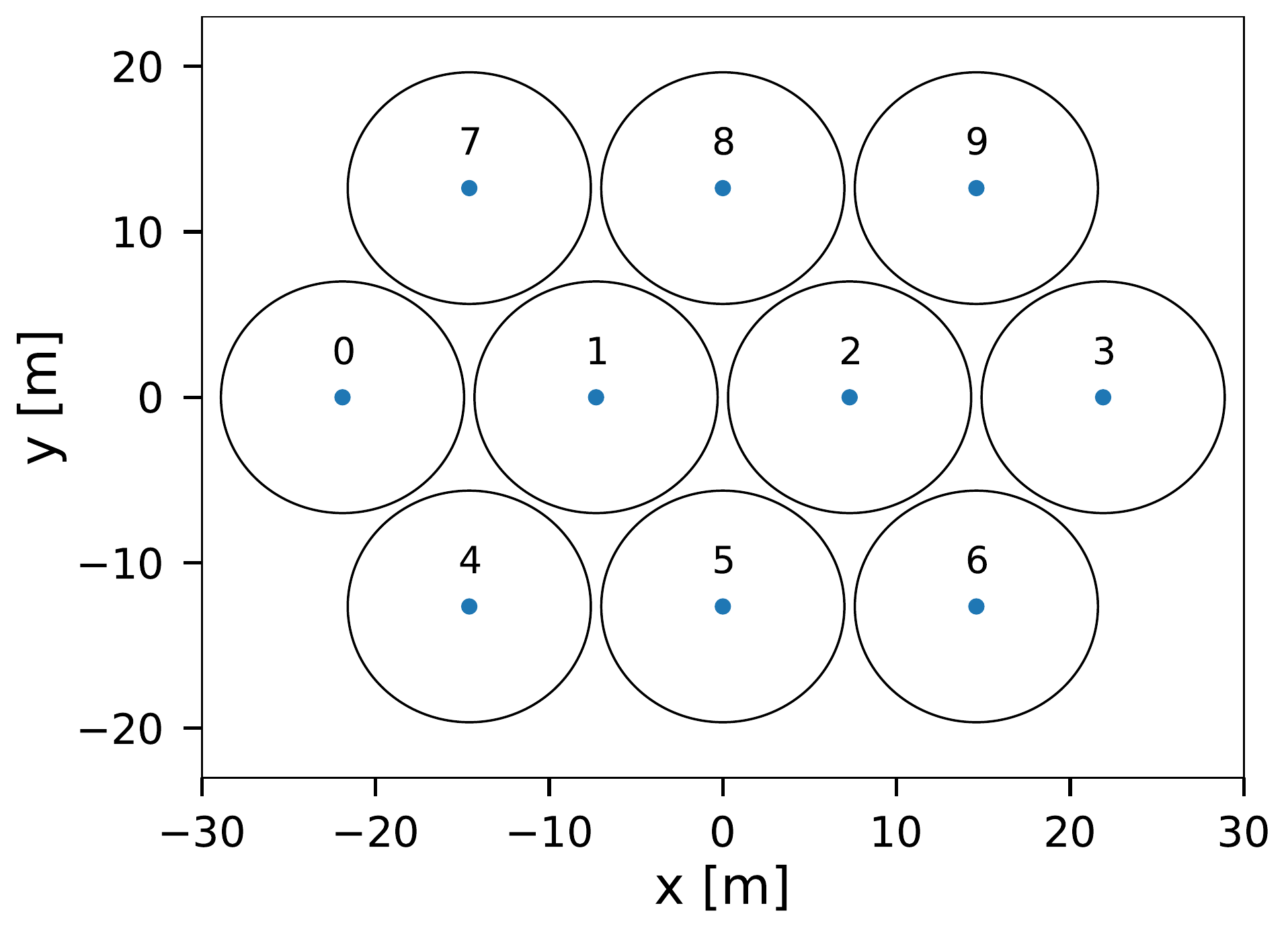}
\caption{The array layout used in our simulations. There are 10 antennas in total, each with diameter 14m and separated by 14.6m.}
\label{fig:fig1}
\end{center}
\end{figure}

\begin{figure*}
\begin{center}
\includegraphics[width=2\columnwidth]{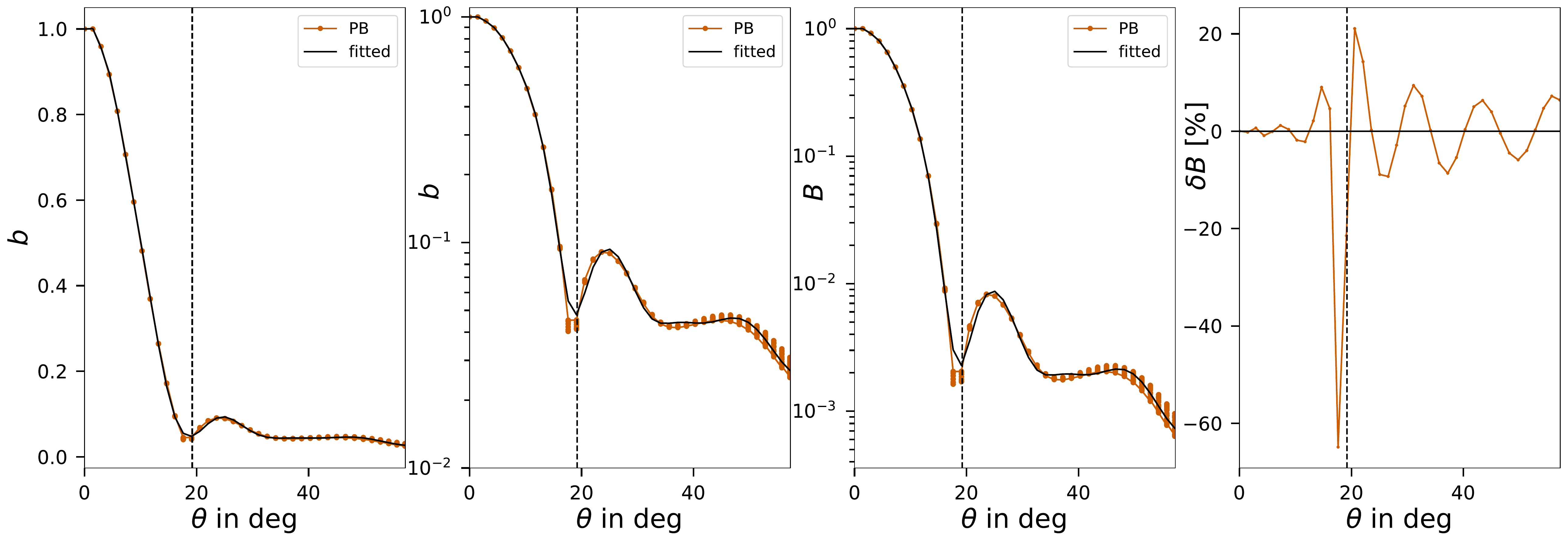}
\caption{{\it (Left):} {Square-root of the EM model primary beam, $b$ (filled circles), and azimuthally-averaged best-fit Chebyshev polynomial with $n_{\rm max}=17$ (black solid line) as a function of zenith angle, $\theta$. Each filled circle denotes a different azimuth angle.}
{\it (2nd from left):} Same as the previous panel, but with a log scale.
{\it (2nd from right):} Comparison of the EM model primary beam and square of best-fit Chebyshev polynomial $(B)$.
{\it (Right):} Fractional (percentage) difference of B between the best-fit Chebyshev polynomial and the EM model primary beam. The fractional deviation is within $10\%$ at all zenith angles except near the first null.}
\label{fig:chebyshev}
\end{center}
\end{figure*}

\section{Array model and types of non-redundancy}
\label{sec:models}

In this section we describe the layout of our model redundant array; a simple analytic model for our fiducial primary beam, based on fits to electromagnetic modelling of HERA antennas; and a series of models for different types of primary beam non-redundancy, based on perturbing the fiducial beam in various ways.

\subsection{Redundant array layout}
\label{sec:array}
A regular, close-packed array layout is generally chosen to ensure a sufficient number of redundant baselines, and to maximise sensitivity on the relatively large angular scales that are targeted by EoR experiments. We consider a hexagonal array layout with similar properties to a segment of HERA, with 10 receivers in our case (see Fig \ref{fig:fig1}). When complete, the full HERA array will comprise 350 dishes, each 14m in diameter, and arranged into three hexagonally-packed sub-arrays with minimum baseline length 14.6m, plus several outrigger antennas to provide longer baselines for imaging. The position error tolerance between antennas is expected to be 2cm or less \citep{2020arXiv200308399D}. As with HERA, we assume that our array operates as a drift scan instrument, pointing at zenith, and located at $-30.7^{\circ}$ latitude. All receivers are assumed to be coplanar and regularly spaced, with no significant height variations or position errors between them.

We chose this array configuration for two reasons. First, our focus in this paper is on primary beam non-redundancies rather than baseline non-redundancies caused by position errors; the latter have been studied elsewhere \citep[e.g.][]{Orosz:2018avj}. Second, 10 close-packed antennas (resulting in 45 baselines in total) is a reasonable minimum to provide several well-populated redundant baseline groups with a few different lengths and orientations while keeping the computational expense of the problem relatively manageable. This allows us to simulate a wider frequency band and LST range for many different types of primary beam non-redundancy. The downside of this choice is that shorter baselines are relatively over-represented compared with the real HERA array. This means that our redundant calibration procedure is more reliant on the baselines that are most sensitive to diffuse emission, although as shown in \citet{Orosz:2018avj} these are less severely impacted by some types of non-redundancy and so may well be up-weighted in a realistic redundant calibration procedure anyway. In any case, we will present results for simulations that include only point sources as well as diffuse emission + point sources, to give some measure of the relative importance of the diffuse emission.

\subsection{Primary beam parametrisation}
\label{sec:beam}
The true (model) visibility for antenna pair $(i, j)$ can be written as
\be
V^\text{true}_{ij} (\nu) = \int_\Omega B_{ij}(\boldsymbol{\theta},\nu) I({\boldsymbol \theta}, \nu) e^{2\pi i {\bf u}_{ij} \cdot\, {\boldsymbol{\theta}}} d^2\Omega
\label{eq2}
\ee
where $I({\boldsymbol\theta}, \nu)$ is the specific intensity in the two dimensional sky plane at position $\boldsymbol\theta$ and frequency $\nu$, ${\bf u}_{ij}$ is the baseline vector, and $B_{ij}$ is the primary beam power pattern corresponding to antennas $i$ and $j$.

\begin{figure*}
\begin{center}
\includegraphics[width=130mm,angle=0]{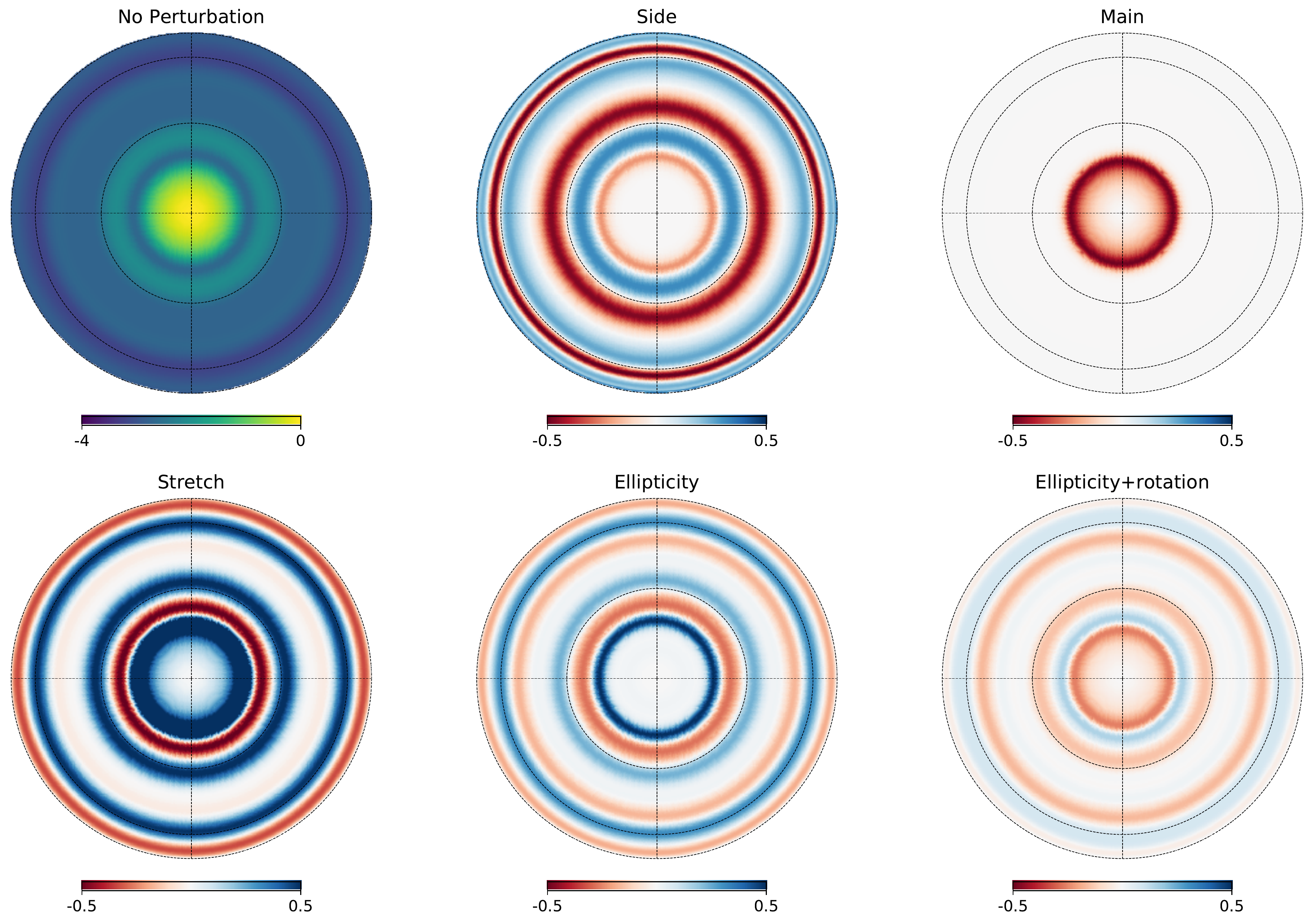}
\caption{Representative {\tt healpix} maps of the perturbed primary beams {in an alt/az coordinate system} for different cases, as explained in Table \ref{table1}. The upper left panel shows the $\log_{10}$ of the power beam for the unperturbed case. All other plots show the ratio of perturbed models to the fiducial model on a $\log_{10}$ scale, clipped to the range $[-0.5, +0.5]$, and for a randomly chosen pair of antennas, $B_{ij}$. Note that these beam patterns will be different for different pairs of antennas.}
\label{fig:beammaps}
\end{center}
\end{figure*}

An important simplification that we make in this paper is to only model the primary beam {\it power} pattern for the (pseudo-) Stokes I polarisation for each antenna. In reality, the power pattern for each baseline is made up of a linear combination of products of electric field patterns from each polarised receiver of each antenna in the pair, encoded by the instrumental Mueller matrix \citep[e.g. see][]{2019ApJ...882...58K}. This accounts for the effects of leakage between different polarisations, which is expected to be a $\sim 1\%$ effect for the HERA Stokes I channel \citep{2019ApJ...882...58K}.

The individual E-field beams per antenna and polarisation are quite complex and do not have even approximate azimuthal symmetry, so it would take a concerted effort (and many parameters) to model them individually. Much of the asymmetry cancels when the Stokes I power beam is formed however, allowing us to construct a reasonably accurate azimuthally-symmetric representation with far fewer parameters. This is the approach we take in what follows -- ignoring polarisation leakage and most beam asymmetry, and approximating the power beam for baseline $(i,j)$ as be $B_{ij}\approx\sqrt{B_i B_j}$, where $B_i$ is the power beam for antenna $i$ for the pseudo-Stokes I polarisation only.

A detailed electromagnetic model for the HERA primary beam was presented in \cite{Fagnoni2021}. While the primary beams of the deployed instrument are likely to deviate from this model due to various effects (e.g. mis-alignments, dish surface imperfections, antenna-antenna couplings), we expect this to be a realistic starting point for our simulations.

Rather than using the simulated beam itself, we fit a purely axisymmetric parametric model of the form
\be
b(\theta, \phi) = \sqrt{B(\theta, \phi)} = \sum_{n=0}^{n_{\rm max}} c_n T_n(x_\nu(\theta)),
\label{eq:3}
\ee
where, for a zenith-pointing drift scan telescope like HERA, $\theta$ is the zenith angle, $\phi$ is the azimuthal angle, and $T_n$ and $c_n$ are the Chebyshev polynomials and their coefficients respectively. Here $x_\nu(\theta)$ is an appropriate frequency-dependent transformation of the angular dependence of the beam that is chosen to make the primary beam pattern easier to fit with a low-order polynomial. We found the transformation $x_\nu(\theta) = 2 \sin(\theta / f_\nu) - 1$ to work well with a variety of polynomial bases, where $f_\nu=(\nu/\nu_0)^\alpha$, with $\nu_0=100~{\rm MHz}$, and $\alpha=-0.69$. We have not introduced any other frequency dependence in the non-redundancy perturbation described in Sect.~\ref{sec:beampert}, which we expect to be a reasonable approximation over the small bandwidth (20 MHz) considered here.

After comparing several different choices of basis expansion, we found that the Chebyshev polynomials with $n_{\rm max} = 17$ match quite well with the electromagnetic model. Fig.~\ref{fig:chebyshev} shows the pixelised beam values in a single frequency channel, compared with the best-fit Chebyshev polynomial. We see that it matches to within $10\%$ at all zenith angles (except two values near the first null, where the primary beam becomes almost zero). The spread in the pixelised beam datapoints due to asymmetry of the beam is also quite small, thus justifying our use of an axisymmetric model.

\subsection{Models of primary beam non-redundancy}
\label{sec:beampert}

\begin{figure*}
\begin{center}
\includegraphics[width=1.9\columnwidth]{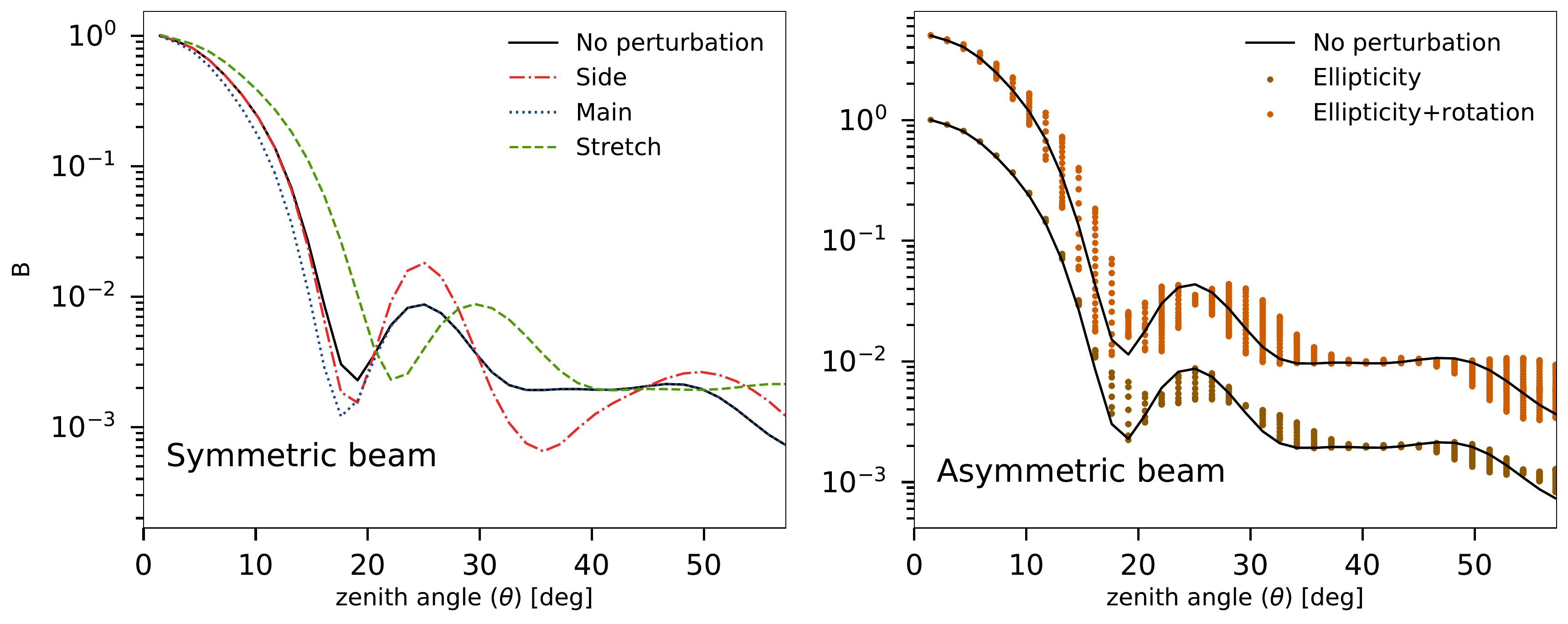}
\caption{The perturbed primary beam as a function of zenith angle for different types of non-redundancy as explained in Table \ref{table1}. The left and right panels show the axisymmetric and asymmetric perturbed  beam respectively. In the right panel, each point, for a fixed zenith angle, is the value of the perturbed beam for different azimuth angle. The solid black line in both panels show the non-perturbed beam for a perfectly redundant array. The ellipticity + rotation curve has been multiplied by a factor of $5$ for ease of comparison with the ellipticity-only case.}
\label{fig:fig4}
\end{center}
\end{figure*}

To model antenna-antenna variations in the primary beam, we introduce a series of different perturbations to the parametric model from the previous section. While it would be possible to construct a more general perturbation scheme, for example using a 2D Zernike polynomial basis, principal component representations, or similar \citep[e.g.][]{2017ExA....44..239B, 2018AJ....156...32E, 2019MNRAS.485.4107I, sekhar19}, our goal here is to minimise the number of additional parameters that must be introduced to the beam model, and to directly connect these parameters to physical effects. Directly perturbing a general polynomial representation of the beam would typically require the coefficients to be adjusted in specific, highly-correlated ways in order to model different effects, since small changes to individual coefficients tend to result in wildly different beam patterns that do not correspond to realistic beam variations. Hence, we have developed a set of specific, physically-motivated perturbations, which are defined as follows (see Table~\ref{table1} and Figs.~\ref{fig:beammaps} and \ref{fig:fig4} for a summary). \\

\noindent \textbf{\textit{Case 1: Sidelobe-only perturbations}} ---
In this case we study the effects of antenna-antenna variations in the sidelobe pattern only, leaving the mainlobe unmodified. In conjunction with Case 2 (which perturbs the mainlobe only), this is intended to allow the relative importance of sidelobe vs mainlobe variations to be compared.

Underlying the model for this case is the assumption that sidelobe variations can be quite complex, potentially shifting the location and depth of nulls in the beam. We use a low-order Fourier series ($N=8$) with randomly-chosen coefficients to modulate the fiducial beam pattern beyond a zenith angle $\theta_{\rm ML}$ that defines the `edge' of the mainlobe. We write the perturbed beam as
\be
\tilde{b}(\theta, \phi) = b(\theta, \phi) \left ( 1 + c_{\rm SL} \sigma_{\rm SL} \Theta(\theta) \sum_m a_m \sin(2 \pi m \theta / L) \right ),
\label{eq4}
\ee
where the period $L = \pi / 2$ corresponds to the angle between zenith and horizon. We normalise the modulation by $c_{\rm SL} = [{{\rm max}(y) - {\rm min}(y)}]^{-1}$, where $y$ is the summation term in Eq.~\ref{eq4}. This rescales the summation term to span $[-1, +1]$ regardless of the chosen values of the coefficients $\{a_m\}$. The parameter $\sigma_{\rm SL}$ then controls the overall amplitude of the modulation. We consider two sub-cases in this paper, with amplitudes $\sigma_{\rm SL}=0.05$ and $0.2$.

We draw the coefficients $\{a_m\}$ randomly from a Gaussian distribution with mean zero and unit variance, using a different set of coefficients for each antenna. The separation between mainlobe and sidelobes is enforced by using a smooth transition of the form
\be
\Theta(\theta) = \frac{1}{2} \left (1 + \tanh \left [ \frac{\theta - \theta_{\rm ML}}{\Delta \theta} \right ] \right ),
\ee
where $\Delta \theta = 3^{\circ}$ determines the sharpness of the transition and $\theta_{\rm ML} \approx 2\, \theta_{\rm FWHM} = 18^{\circ}$ sets its location. This prevents sharp artifacts from appearing in the beam model. An example realisation of this kind of perturbation is shown in Fig.~\ref{fig:beammaps} and Fig.~\ref{fig:fig4}. \\

\begin{table}
  \centering
  \begin{tabular}{c|ll}
   & {\bf Perturbation type} & {\bf Perturbation level} \\
  \hline
  {\bf Case 1} & Sidelobe & (a) $\sigma_{\rm SL} = 0.05, 0.2$ \\
  & & (b) $\sigma_{\rm SL} = 0.2$, $\sigma_{\rm freq} = 0.1$  \\
  \hline
  {\bf Case 2} & Mainlobe & $\sigma_{\rm ML} = 0.01, 0.02$ \\
  \hline
  {\bf Case 3} & Stretched beam  & (a) Gaussian, $\sigma_m = 0.01, 0.02$\\ 
  & & (b) Uniform, $\Delta_m = 0.02$ \\
  & & (c) Outlier antenna 2 \\
  & & (d) Outlier antenna 7  \\
  \hline
  {\bf Case 4} & (a) Ellipticity &  $\sigma_{m,x/y} = 0.01, 0.02$ \\ 
  & (b) Ellipt. + rotation &  $\sigma_{m,x/y} = 0.01, 0.02$ and\\
  &  & Rotation $\alpha \sim {\rm Uniform}[0^{\circ},360^{\circ}]$ \\
  & (c) Fixed ellipt., rotation: & (i) $\alpha \sim {\rm Uniform}[0^{\circ},360^{\circ}]$  \\
  & & (ii) $\alpha \sim {\rm Gauss.}(\mu=0, \sigma=10^{\circ}$) \\
  \hline
  \end{tabular}
  \caption{The different types of primary beam non-redundancy considered in this study, along with the parameters used to define the corresponding primary beam perturbations.}
  \label{table1}
\end{table}

\begin{figure*}
\begin{center}
\includegraphics[width=2.\columnwidth,angle=0]{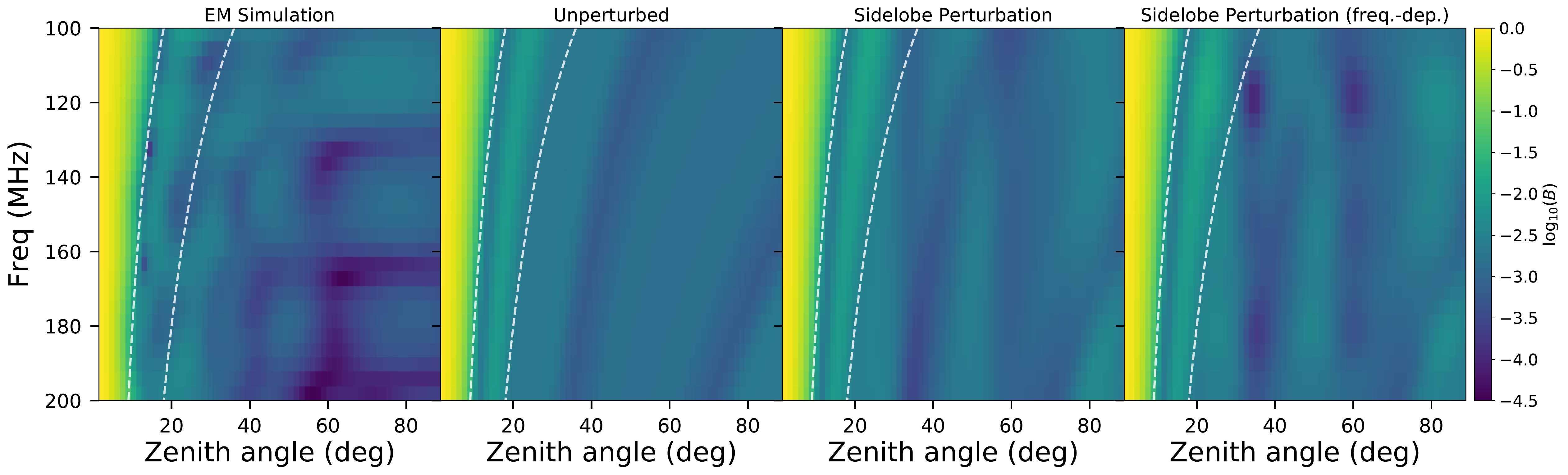}
\caption{{Waterfall (frequency-zenith angle) plots of different models of the primary beam, $B(\theta, \nu)$. {\it (Left):} Azimuthal average of the full EM-simulation beam of \citet{Fagnoni2021}. {\it (2nd from left):} Unperturbed Chebyshev polynomial fit to the EM-simulated beam (Eq.~\ref{eq:3}). {\it (2nd from right):} Chebyshev model plus a sidelobe perturbation with $\sigma_{\rm SL} = 0.4$. {\it (Right):} Chebyshev model with the same sidelobe perturbation, but now with a large additional frequency modulation with $\sigma_{\rm freq} = 0.99$. The two white dashed lines show the positions of the first and second nulls in the Chebyshev model (at $\sim 18^{\circ}$ and $\sim 36^{\circ}$ respectively at 100 MHz).}}
\label{fig:polyfreq}
\end{center}
\end{figure*}

\noindent\textbf{\textit{Case 2: Mainlobe-only perturbations}} ---
The mainlobe of the HERA primary beam approximately follows a Gaussian. We perturb the mainlobe by subtracting a Gaussian function with the same width as the mainlobe of the fiducial model, and then adding another Gaussian with a modified width. The former Gaussian has width (standard deviation) $\theta_{\rm ML}$, while the latter has $\gamma\theta_{\rm ML}$, where $\gamma$ controls the width of the perturbed main lobe. When $\gamma$ is greater than 1, the width of the perturbed primary beam is larger compared with the fiducial model (and vice versa). The mathematical expression for the perturbed primary beam is
\bea
\tilde{b}(\theta, \phi) &=& b(\theta, \phi) + (1 - \Theta(\theta)) \left [ q(\theta; \gamma) - q(\theta; \gamma=1) \right ] \\
q(\theta; \gamma) &=& \exp \left (-\frac{1}{2}\frac{\theta^2}{\gamma^2 \theta_{\rm ML}^2} \right ).
\eea
We draw $\gamma$, the width perturbation parameter, from a Gaussian random distribution with mean unity and standard deviation $\sigma_{\rm ML}$. Since this can be larger or smaller than unity, the perturbed mainlobe can be larger or smaller than the fiducial one. We consider two different cases here, with $\sigma_{\rm ML}=0.01$ and $0.02$.\\

\noindent \textbf{\textit{Case 3: Stretching the primary beam}} ---
This case models changes in the overall angular size of the beams from antenna to antenna. Similar kinds of non-redundancy might arise if the height of the receiver above the dish varies slightly between antennas for example. For simplicity, the beam pattern is left unchanged, preserving the structure of the mainlobe and sidelobes, except for an overall stretching factor that varies between dishes. This is achieved by performing a remapping of the zenith angle, $\theta \to \theta / m$,
where $m$ is the stretch factor. We consider three different cases:
\begin{enumerate}[label=(\alph*),labelwidth=2em,leftmargin=\parindent]
 \item For each antenna, $m$ is drawn from a Gaussian distribution with mean unity and standard deviation $\sigma_m=0.01$ or $0.02$.

 \item For each antenna, $m$ is drawn from a Uniform distribution between $[-0.02, +0.02]$ (roughly comparable in width to the $\sigma_m = 0.01$ case above).

  \item A perfectly redundant array with $m=1$ for all antennas except a single outlier, which has a $10\%$ stretch factor applied. We consider cases with the outlier antenna in the middle of the array (Ant.~2), and on the outskirts (Ant.~7); see Fig \ref{fig:fig1}.
\end{enumerate}
Because each visibility depends on the product of the square roots of the individual antennas' power beams, this type of non-redundancy can generate complex deviations from the fiducial model, as can be seen from Fig.~\ref{fig:beammaps}. In essence, the primary beams of each antenna in the pair modulate one another, generating substantial additional structure, especially in the sidelobes. \\

\noindent\textbf{\textit{Case 4: Ellipticity and rotation of the primary beam}} ---
While the basic beam model is axisymmetric, we also allow perturbations in ellipticity and rotation to model beam squint and feed rotation effects. We do this by transforming the axisymmetric beam through a simple coordinate remapping,
\bea
&&x = \theta \cos\phi;~~~ y = \theta\sin\phi \nonumber\\
&&x \to (x \cos \alpha - y \sin \alpha) / m_x \nonumber\\
&&y \to (x \sin \alpha + y \cos \alpha) / m_y \nonumber,
\eea
where $m_x$ and $m_y$ are stretch factors in the E-W and N-S directions and $\alpha$ is the rotation angle of the ellipse semi-major axis away from the E-W direction. When $m_x\neq m_y$, ellipticity is generated in the perturbed beam.

The shape of the perturbed primary beams for different cases are shown in Fig.~\ref{fig:beammaps} and Fig.~\ref{fig:fig4}. The former shows a {\tt healpix} map \citep{gorski05} of the perturbed beam {in an alt/az coordinate system}, in which most of the visible structure can be attributed to an axisymmetric perturbation due to the stretch factors deviating from unity. Fig.~\ref{fig:fig4} more clearly shows the asymmetry as a function of zenith angle.

We consider three different cases for ellipticity and rotation:
\begin{enumerate}[label=(\alph*),labelwidth=2em,leftmargin=\parindent]
  \item We draw two different Gaussian random numbers for $m_x$ and $m_y$ for each antenna, with $\sigma_m=0.01$ and $0.02$. This results in either N-S or E-W-aligned elliptical beams, with no rotation.

  \item Same as (a), but also drawing a random rotation angle $\alpha$ for each antenna from a Uniform distribution over $[0^{\circ}, 360^{\circ}]$.

  \item We fix the ellipticity $(m_x=1.02, m_y=0.98)$, but draw a rotation angle for each antenna from either a Uniform distribution in the range $[0^{\circ}, 360^{\circ}]$, or a Gaussian distribution with mean $0^{\circ}$ and standard deviation $10^{\circ}$.
\end{enumerate}
{The functional forms of the different types of non-redundancy discussed so far do not have an explicit frequency dependence of their own. The perturbed beams can still have differing, non-trivial frequency structures, however. This is because most of the types of non-redundancy are implemented as modulations of the unperturbed beam model, which has a simple power-law frequency scaling as described by Eq.~\ref{eq:3}, and so the beam shape as a function of frequency does differ between antennas.

Fig.~\ref{fig:polyfreq} compares the resulting (mildly) frequency-dependent structure for an example of the sidelobe perturbation model with the unperturbed model and the full EM-simulation model of \citet{Fagnoni2021}. While the zenith angle-dependent modulation clearly dominates the perturbed sidelobe structure, the underlying frequency-dependence of the unperturbed model is still visible, and will be affected in different ways for different realisations of the perturbations (i.e. between different antennas).

For completeness, we also include a test case with an explicitly frequency-dependent sidelobe model, which is listed in Table~\ref{table1} as Case 1(b), and an extreme example of which is shown in the right most panel of Fig.~\ref{fig:polyfreq}. We implement the frequency dependence by promoting the sidelobe perturbation amplitude parameter to a function of frequency,
\be \label{eq:slfreq}
c_{\rm SL} \to c_{\rm SL} \sigma_{\rm freq} \sum_n 
  c_n \sin \left (\frac{2\pi \nu}{100~{\rm MHz}} \right ) 
+ d_n \cos \left ( \frac{2\pi \nu}{100~{\rm MHz}} \right ),
\ee
where random coefficients $c_n$ and $d_n$ are chosen for a low-order Fourier series using a similar method as in Eq.~\ref{eq4}, with $\sigma_{\rm freq}$ now controlling the size of the perturbation. For the extreme case of $\sigma_{\rm freq} = 0.99$ shown in Fig.~\ref{fig:polyfreq}, the additional frequency-dependent modulation gives rise to features such as an increased number of sidelobes at around 120 MHz. We will study the less extreme case of $\sigma_{\rm freq} = 0.1$ in Sect.~\ref{sec:freqsl}.}


\begin{figure*}
\begin{center}
\includegraphics[width=2\columnwidth,angle=0]{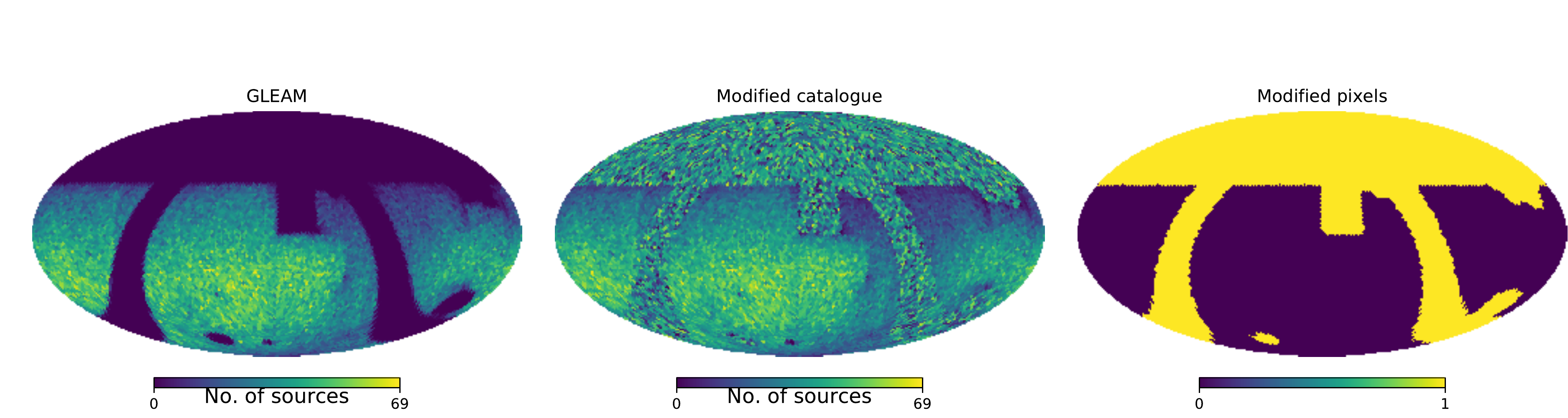}
\caption{{\it (Left):} The distribution of the number of sources per pixel in a {\tt healpix} map with {\tt nside = 32} from the GLEAM catalogue. {\it (Middle):} Same as the left panel but for a modified GLEAM catalogue where the blank pixels are replaced with a random pixel from the filled regions. {\it (Right):} The modified pixels are shown with value 1. {An Equatorial coordinate system and Mollweide projection have been used.}}
\label{fig:fig5}
\end{center}
\end{figure*}

\section{Simulations}
\label{sec:sims}

In this section we describe our suite of visibility simulations for the array layout and set of primary beam models described in the previous section. For all simulations, we use a bandwidth of $100-120$ MHz with 120 frequency channels, and a total observation time of about 6.7 h with an integration time per sample of 40 sec. Our simulations cover the LST range $9.2 - 15.8$ h. This range is chosen to be almost disjoint with recent seasons of HERA data (covering $\sim 0 - 11.5$ h), as the modelling presented here is not yet suitable for a direct comparison. The whole analysis presented here is for the pseudo-Stokes I polarisation only.

We use the {\tt hera\_sim}\footnote{\url{https://github.com/HERA-Team/hera_sim/}} and {\tt healvis} \citep{2019ascl.soft07002L} packages to perform the simulations for point sources and diffuse emission respectively. Both implement the visibility equation (Eq.~\ref{eq2}), but with different ways of converting between sky and antenna coordinates and different ways of modelling the beams. We modified the {\tt hera\_sim} package to use a fast approximation to the angle conversions in the {\tt astropy} package \citep{2018AJ....156..123A}, which produces results that accurately match those from the {\tt pyuvsim} high-precision reference simulator \citep{2019JOSS....4.1234L} at significantly less computational expense. We also modified the packages to evaluate the primary beams directly, using the analytic model to avoid any additional interpolation or pixelisation steps.

\subsection{Point sources}
\label{sec:ptsrc}

The target EoR signal is several orders of magnitude fainter than typical foreground emission, and so we expect foregrounds to be the dominant contributor to any systematic errors in the calibration solutions due to non-redundancies. We therefore neglect the EoR component, and instead generate our initial sky model from point source foregrounds only. We base our sky model on the GLEAM catalogue \citep{hurley17}, as shown in Fig.~\ref{fig:fig5}. GLEAM contains 307,455 sources over a total sky area of 24,831 deg$^2$, with variable depth and completeness (representative completeness of 50\% complete at 55 mJy). Each source has flux density measurements in 20 sub-bands in the range $72-231$ MHz, making it well-matched to our simulated frequency range of $100-120$ MHz.

The GLEAM catalogue excludes the region north of $+30^{\circ}$ declination, Galactic latitudes within $10^{\circ}$ of the Galactic plane, and a handful of localised areas such as the Magellanic Clouds. To ensure a realistic sky brightness distribution even in the far sidelobes of the primary beam for the entire LST range of our simulations, we fill in the excluded regions in the catalogue with sources drawn from elsewhere in the catalogue. We assign all the sources to pixels within a {\tt healpix} pixelisation with {\tt nside = 32}. We then choose a random pixel from the observed region (see the left panel of Fig.~\ref{fig:fig5}) and duplicate the sources within that pixel in an empty pixel elsewhere in the map. We repeat until all empty pixels are filled, randomly selecting a new observed pixel each time to ensure a fair sampling of the catalogue.

The modified source distribution is shown in the middle panel of Fig.~\ref{fig:fig5}, with the right panel showing the pixels that have been populated using this filling process. It can be seen from the source distribution map that the filled pixels do not reproduce the large-scale clustering properties of the regions that were included in the actual catalogue, and that artifacts remain towards the edges of the survey region due to the diminishing depth of the real survey. We make no attempt to correct for these issues, as we expect the brightest sources to be the dominant contributors to non-redundancy effects, and these are treated separately (see below). The overall luminosity function of these fainter sources across the whole sky is consistent with that within the GLEAM survey, and there are no unrealistic empty regions in the sky brightness distribution, and so we consider this treatment sufficiently realistic for our purposes.

\begin{figure}
\begin{center}
\includegraphics[width=\columnwidth,angle=0]{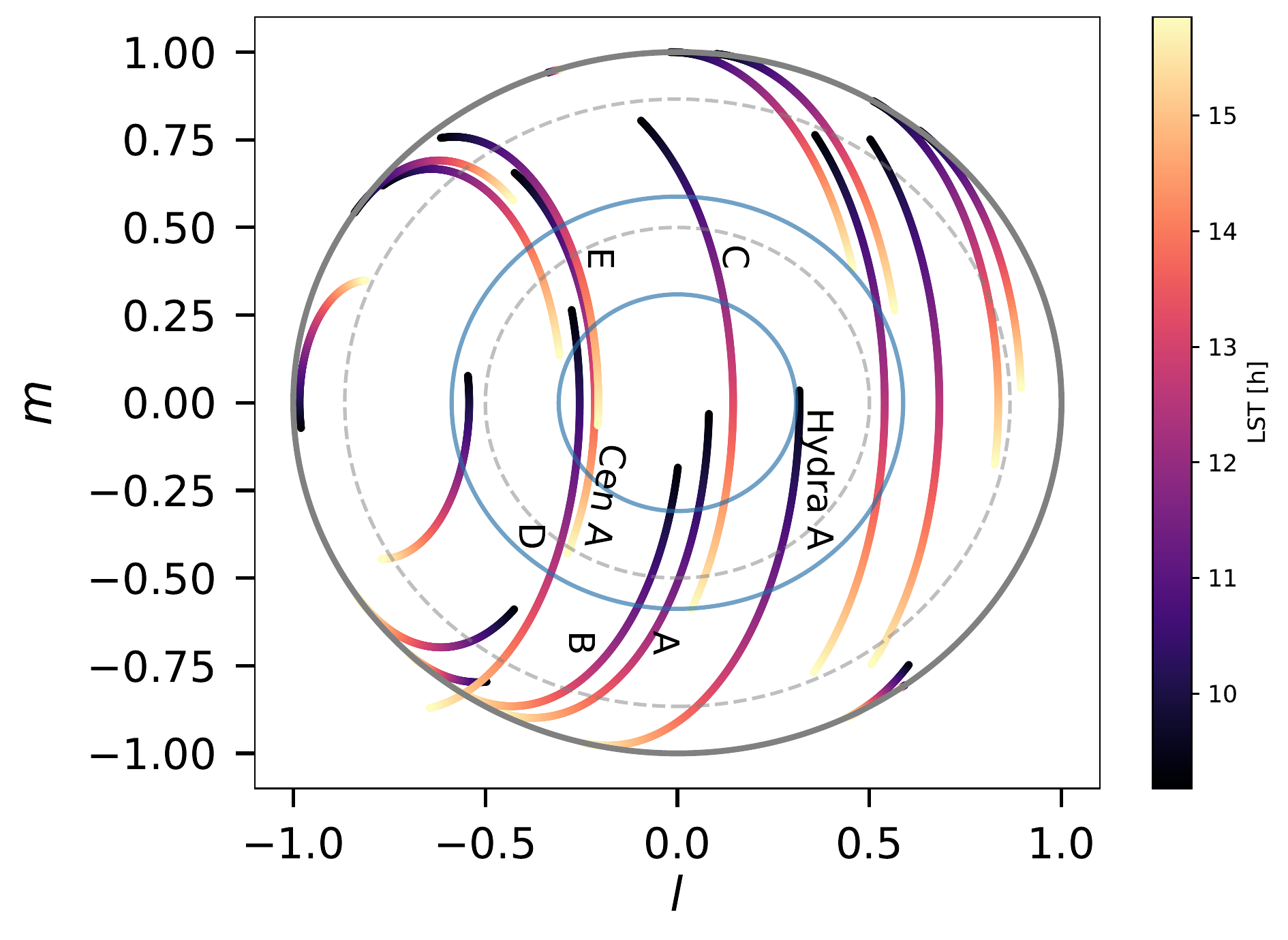}
\caption{Tracks of the brightest 20 point sources through the primary beam as a function of LST. Tracks for a few other important sources (A to E) are also shown; see Table~\ref{table2} for a listing. Note that the tracks of D and E have been shifted slightly for clarity. The blue circles denote the first (inner) and second (outer) nulls in the primary beam, at around 18$^\circ$ and 36$^\circ$ respectively, while the grey dashed circles are placed at 30$^\circ$ and 60$^\circ$.}
\label{fig:tracks}
\end{center}
\end{figure}

We include the brightest sources that were peeled from GLEAM catalogue itself (see Table 2 of \citealt{hurley17}), {as well as {\it Fornax A}, which is not present in that table.} Fig.~\ref{fig:tracks} shows the tracks as a function of LST of the 20 brightest sources for the whole observation through the primary beam of our simulated array. Five other bright sources with fluxes $\sim 30-60$ Jy that pass through the beam are also included, as summarised in Table~\ref{table2}. Not all of these bright sources rise above the horizon within the simulated 6.7 h observing time. The blue circles in this figure denote the first (inner) and second (outer) nulls in the primary beam pattern, at around 18$^\circ$ and $36^\circ$ respectively (c.f. Fig.~\ref{fig:chebyshev}). We see that Hydra~A (545 Jy at 100 MHz) is present near the first null at the beginning of the observation, while Cen~A (1937 Jy at 100 MHz) is present within the mainlobe from around 13 h.

\begin{figure}
\begin{center}
\includegraphics[width=\columnwidth,angle=0]{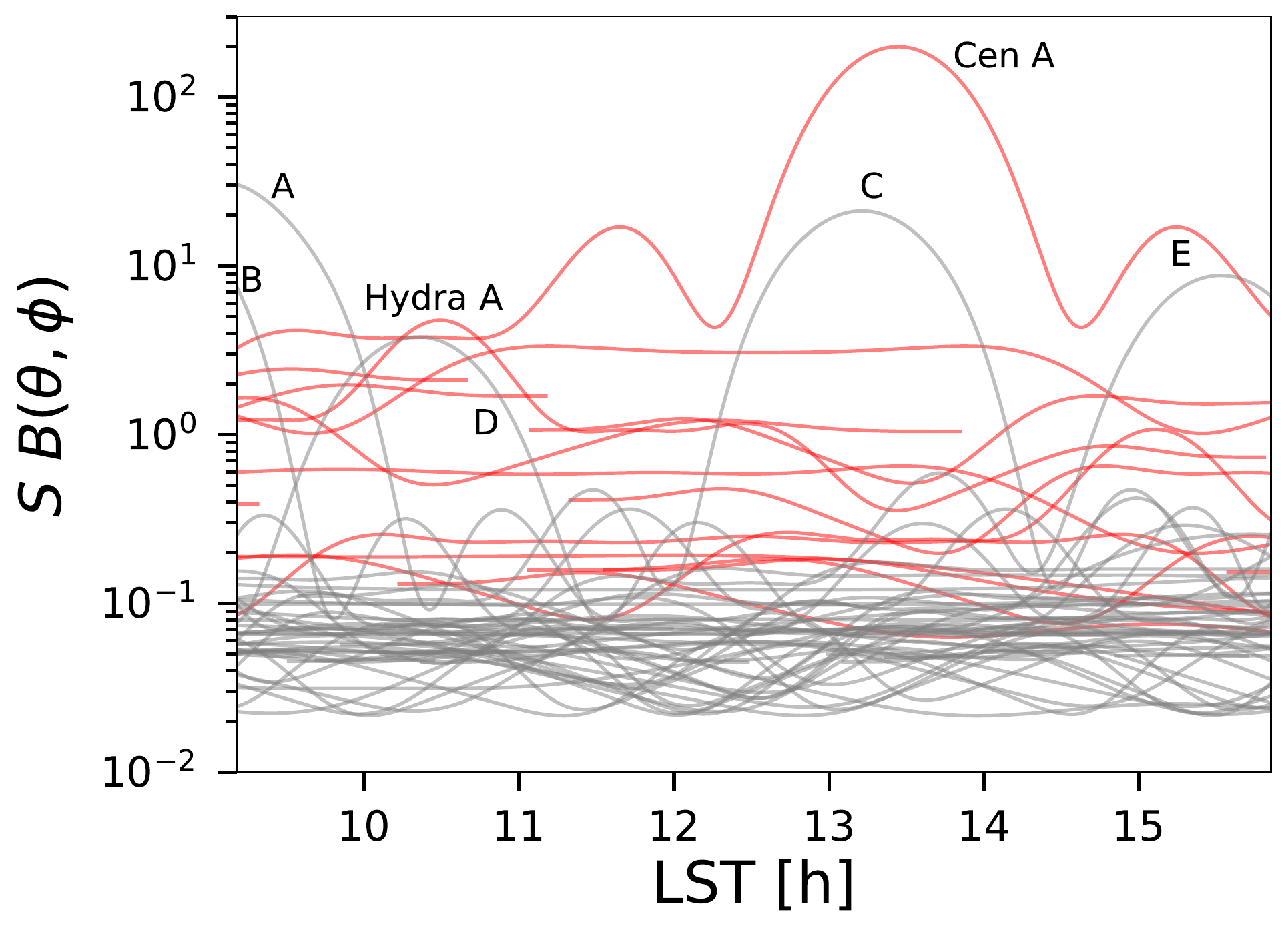}
\caption{The effective flux (flux $S$ times primary beam $B$) as a function of LST for the 100 brightest sources. The red lines are for top 20 brightest sources (c.f. Fig.~\ref{fig:tracks}), while the grey lines are for remaining 80 sources. Not all of the sources are present in the plot due to the limited LST range considered here. Sources that were found to have an important effect on the gain solutions are labelled (see Table~\ref{table2}).}
\label{fig:pbflux}
\end{center}
\end{figure}

In Fig.~\ref{fig:pbflux}, we show the the flux of the 100 brightest sources as a function of LST after multiplying with the analytic primary beam (square of Eq.~\ref{eq:3}). The red lines are for the top 20 brightest sources, while the grey lines are for the next 80 brightest sources. The brightest sources mostly dominate the effective flux level for the entire LST range, except at LSTs of around $9-10$ h where sources A and B (fluxes of $\sim 41$ Jy and $36$ Jy respectively) form the dominant contribution. Sources C and E (fluxes of $\sim 54$ Jy and $68$ Jy respectively) also contribute significantly at around 13 h and 15.5 h due to their position within the mainlobe at those times, although Cen~A is still by far the dominant source from around 11 h onwards.

As mentioned previously, not all of the 100 brightest sources are present in the plot due to the limited LST range considered here. The effective flux (flux times primary beam) of each source changes with LST in a characteristic way depending on how exactly it transits through the beam pattern. Most of the bright sources show an oscillatory pattern with LST as they pass through different sidelobes and nulls, while a handful of sources transit directly through the mainlobe and so are particularly conducive to empirically mapping the beam pattern \citep{2020ApJ...897....5N}.

Note that we do not include any bright extended sources in our sky model, other than the Galactic diffuse emission (see below). Given the comparatively low angular resolution of our simulated array, we expect extended sources to be adequately modelled as point sources for our purposes.

\subsection{Diffuse emission}
\label{sec:diffuse}

In addition to point sources, we also include diffuse emission in some of our simulations. This is simulated separately using the {\tt healvis} package, with the same simulation properties and primary beam models as for the point sources. The resulting visibilities are then added to the point source simulations, followed by noise and gain fluctuations (see below). Note that the addition of diffuse emission significantly alters the sky temperature, and therefore increases the noise rms of the simulations compared with the point source-only case.

We use the Global Sky Model \citep{gsm} as our model of the diffuse Galactic radio foregrounds, as implemented by the {\tt PyGSM} package \citep{pygsm}. GSM is based on a three-component principal component analysis fit of a large set of multi-frequency datasets, spanning 10 MHz to 94 GHz. At low frequencies, these components roughly correspond to the Galactic synchrotron and free-free emission that are the dominant contributors to the sky temperature around 100 MHz. The model comes in the form of {\tt healpix} maps as a function of frequency, which we pass to {\tt healvis} with a map resolution of {\tt nside=64}. We also performed a test simulation with {\tt nside=128}, finding little change in the model visibilities for a significant increase in computational expense, hence our choice of the lower resolution. Note that the diffuse emission simulations represent the dominant computational cost in this study.

\begin{table}
  \centering
  \begin{tabular}{c|llr}
  \hline
   {\bf Name} & {\bf RA [$^\circ$]} & {\bf Dec [$^\circ$]} & {\bf Flux [Jy]} \\
  \hline
  {\bf J090147-255516 (A)} & $135.447$ & $-25.921$ &  40.9\\
  \hline
  {\bf B} & $125.2$ & $-30.0$ &  36.2\\
  \hline
  {\bf J131139-221640 (C)} & $197.914$ & $-22.277$ & 53.7 \\
  \hline
  {\bf J102003-425130 (D)} & $155.015$ & $-42.858$  & 34.3 \\
  \hline
  {\bf J153014-423146 (E)} & $232.558$ & $-42.529$  & 67.5 \\
  \hline
  {\bf Centaurus A} & $201.3$ & $-43.0$  & 1937.4 \\
  \hline
  {\bf Hydra A} & $139.5$ & $-12.1$ & 544.7 \\
  \hline
  \end{tabular}
  \caption{The position and flux (at 100 MHz) of the dominant sources in our simulations. The common name or GLEAM identifier is given for each source; source B is actually a randomly-drawn source used to fill in a blank region, as discussed in Sect.~\ref{sec:ptsrc}. The tracks and the effective (beam-modulated) fluxes of these sources are shown in Fig.~\ref{fig:tracks} and \ref{fig:pbflux} respectively.}
  \label{table2}
\end{table}

\subsection{Gains and noise}
\label{sec:simgain}

In addition to the sky and beam models, we also introduce a simple model for the bandpass, by generating complex gains for each antenna that vary smoothly in frequency (but not in time). The gains are generated from a complex sine series,
\be
g_i(\nu,t) = \frac{(1+i)}{\sqrt{2}} + f \sum^{n_{{\rm modes}}-1}_{m=0} (a_m + ib_m) \frac{\sin(\pi m\nu)}{m+1}
\ee
where ${a_m,b_m}$ are Gaussian random variates with mean zero and unit variance, and $n_{\rm modes}=8$ is the number of sinusoidal modes used to generate the gain variation along the frequency. Different coefficients are drawn for every antenna. The value of $f$ determines the amplitude of the bandpass fluctuations, which we set to $f=0.1$ in all simulations.

For the noise part, we add uncorrelated Gaussian random noise to each frequency channel and time sample according to
\be
n_{ij}(\nu, t) = \sigma_{ij} \frac{(a+ib)}{\sqrt{2}}, 
\ee
where $\sigma^2_{ij}$ is the desired noise variance and $a$ and $b$ are unit Gaussian random variates. The noise variance is modelled using the simulated autocorrelation visibilities, according to
\be
\sigma_{ij}^2 = {\frac{V_{ii}V_{jj}}{\Delta t \Delta \nu}},
\ee
where $\Delta t=40~{\rm sec}$ is the time resolution and $\Delta \nu=166~{\rm kHz}$. This ensures that the noise level tracks the sky temperature, which we assume to dominate the system temperature at these frequencies.

Note that we do not model more complex instrumental effects such as RFI, polarisation leakage, or cross-talk in these simulations.

Finally, we apply the gains and noise to the true simulated visibilities according to Eq.~\ref{eq:gains}, and store the results in the standard {\tt UVData} format used by the HERA analysis pipeline. Our simulation pipeline is available from \url{https://github.com/philbull/non-redundant-pipeline/}, and includes configuration files that can be used to regenerate the simulated data for each type of primary beam non-redundancy.

\section{Analysis pipeline}
\label{sec:analysis}

In this section, we describe a simplified analysis pipeline that we apply to each simulated dataset, based on the HERA data analysis and power spectrum estimation pipelines. Our simplified pipeline consists of a redundant calibration step to solve for the gains (Sect.~\ref{sec:redcal}), followed by an artificial absolute calibration to fix degeneracies in the gain solutions, then an optional coherent averaging of the visibilities, and finally delay spectrum estimation (Sect.~\ref{sec:pspec}).

We compare two different approaches to estimating the delay power spectrum:
\begin{enumerate}[label=(\alph*),labelwidth=2em,leftmargin=\parindent]
  \item {\it Incoherent averaging}: Estimate the delay spectra for pairs of baselines within each redundant group, and then average the resulting spectra to form an incoherently-averaged power spectrum.
  
  \item {\it Coherent averaging}: Average together the visibilities for all pairs of baselines within each redundant group and then estimate the delay spectrum for the average to form a coherently-averaged power spectrum.
\end{enumerate}
For a set of delay-transformed visibilities $\{\tilde{V}_a(\tau, t)\}$ within a redundant baseline group, where $a$ labels the baseline, the incoherently-averaged auto-spectra and coherently-averaged (auto- and cross-) delay spectra are given by, respectively,
\bea
  P_{\rm inco}(\tau, t) &=& \frac{1}{N_{\rm bl}}\sum_a \tilde{V}_a(t)\, \tilde{V}_a^*(t + \Delta t) \label{eq:Pinco}\\
  P_{\rm co}(\tau, t) &=& \frac{1}{N_{\rm bl}^2} \left ( \sum_a \tilde{V}_a(t) \right ) \left (\sum_b \tilde{V}_b^*(t+\Delta t) \right ), \label{eq:Pco}
\eea
where $N_{\rm bl}$ is the number of baselines within the redundant group, $t$ labels the LST bin of the visibility observation, and $\Delta t$ is the LST bin width, so that the spectra are calculated from neighbouring time samples to avoid a noise bias.

The two approaches should give equivalent results in the ideal case, up to a difference in their noise levels. Cancellations due to phase errors in the gain solutions, caused by non-redundancy, can cause a systematic decoherence effect in the coherently-averaged visibilities however, leading to the possibility of signal loss. In contrast, the incoherently-averaged auto-spectra should not be susceptible to this form of signal loss (HERA Collaboration 2021, {\it in prep.}), as the phase error cancels exactly in the $\Delta t = 0$ case, and should remain small (for auto-baseline spectra) when $\Delta t$ is much smaller than the beam-crossing time. We will study this effect in Sect.~\ref{sec:results}.

\subsection{Redundant gain calibration}
\label{sec:redcal}

As we will show, the effect of non-redundancy is to induce spurious spectral and temporal structure in the antenna gain solutions. How this arises depends strongly on the calibration method being used; an algorithm that assumes perfect redundancy of baselines or makes strong assumptions about the primary beam of each antenna is likely to be more sensitive to different types of non-redundancy for example.

We use the redundant calibration method described in \cite{2020arXiv200308399D} implemented in the {\tt hera\_cal} package\footnote{\url{https://github.com/HERA-Team/hera_cal/}} to derive a redundant calibration from each of our simulations. The stages of this method are as follows:
\begin{enumerate}[label=(\alph*),labelwidth=2em,leftmargin=\parindent]
  \item Approximate solutions for the phase of the gains are first derived using an iterative calibration that assumes perfect array redundancy ({\tt firstcal});
  \item A single iteration of the {\tt logcal} algorithm is applied, based on taking the logarithm of Eq.~\ref{eq:gains} to linearise it, and then solving for the free gain and model parameters;
  \item Repeated iterations of the {\tt omnical} algorithm, which jointly solves for the complex gains and redundant model visibilities up to a set of degeneracies.
\end{enumerate}
This procedure estimates the gains for each frequency channel and time sample independently, and we do not apply any smoothing to the solutions to account for the fact that they are expected to have some degree of smoothness.

Because the data break the assumption of baseline redundancy that the algorithm is based on, we expect the {\tt redcal} solver to absorb some of the deviations from perfect redundancy into the gain and visibility model solutions, while leaving an additional unmodelled residual per baseline that can't be fit by a redundant calibration model. The way that these errors manifest is sensitive to the details of the redundant calibration process, such as which antennas/baselines are included in the fits or not, and how convergence criteria (e.g. $\chi^2$ fitting statistics) are handled. For example, previous work on non-redundancy due to antenna position errors by \cite{Orosz:2018avj} found that gain errors could be reduced by excluding the longest baselines from the {\tt redcal} calibration process, since they are more strongly affected by chromatic errors introduced by the non-redundancy. Given the relatively small size of our simulated array, we have effectively incorporated this recommendation automatically in this paper.

After {\tt redcal} has run, we find a small number of outliers in our gain solutions, where the fitting procedure seems to have failed. The outliers tend to affect all frequency channels for the duration of a few time samples, and are characterised by a clear discontinuity in the otherwise smoothly-varying phase of the gain solutions (with no such discontinuity visible in the amplitude, and a normal $\chi^2$ statistic for the affected solutions reported by the algorithm). This issue is thought to be peculiar to relatively small arrays, where there are fewer constraints on each calibration degree of freedom, making instabilities of this kind more likely ({\it J.~Dillon, priv. comm.}). To address this issue, we first identified the outliers in the gain solutions across the time and frequency axes in turn, for all antennas, by using a median absolute deviation (MAD) filter. Any datapoint $d$ that satisfies the condition $(d - {\rm median}(d)) / {\rm MAD}(d) > 20$ after the gain calibration has been applied is flagged as an outlier (where the median is calculated over the time and frequency axes in turn). We then replace the outlier gain solutions with the mean of their nearest unflagged neighbour points. Since there are only a few outliers for any given dataset, we expect this replacement to only have a small effect on the final delay spectra, whereas leaving in the outliers would have caused a substantial amount of ringing. A handful of smaller outliers do remain following this process in some cases, which we have not addressed. They can be seen as localised (in time) spikes in some of our results (e.g. see the Outlier Ant2 panel in Fig~\ref{fig:gaindiff}). We do not expect them to change our conclusions however, and note that in real-world scenarios, these would likely either not arise (due to the array being larger) or would be removed by gain smoothing.

Following redundant calibration, a final absolute calibration step, based on a sky model, is typically applied to fix the small number of degenerate parameters that cannot be determined through redundant calibration. These include an overall amplitude and phase offset, and tip-tilt parameters related to the overall orientation of the array \citep{2020ApJ...890..122K}. We must also fix these degrees of freedom in our analysis, but wish to do so without adding the complication of a sky model-based absolute calibration step, which may introduce additional gain errors beyond the kind we are studying here. Instead, we fix the degenerate degrees of freedom to their values from the true, simulated gains, using no other information from the true gains to inform the gain solutions.

As a final step, we apply the redundantly-calibrated and degeneracy-fixed gain solutions to the simulated data in order to recover an estimate of the true visibilities, which we refer to as the calibrated visibilities, $V_{ij}^{\rm cal}$. We use the calibrated visibilities for each baseline in our subsequent analysis, rather than the redundant model visibilities that are also output by the {\tt redcal} algorithm.

\subsection{Power spectrum estimation}
\label{sec:pspec}

We use the {\tt hera\_pspec}\footnote{\url{https://github.com/HERA-Team/hera_pspec}} package to estimate the power spectrum of the visibilities. {\tt hera\_pspec} uses the optimal quadratic estimator (OQE) formalism under the delay approximation. The delay-transformed visibility is the Fourier transform along the frequency direction, and can be written as \citep{Parsons:2012qh}
\be
\tilde{V}_{ij}(\tau)=\int d\nu e^{2\pi i\nu\tau} W(\nu) V_{ij}(\nu),
\ee
where the delay $\tau = \vec{b}_{ij} \cdot \hat{n} / c$ is defined for the baseline vector $\vec{b}_{ij}$ between two antennas, and a direction on the sky $\hat{n}$. The taper, $W(\nu)$, is chosen as a Blackman-Harris window, which is applied in the frequency domain to make the visibility periodic within the bandwidth. This has the effect of reducing ringing, while effectively correlating neighbouring Fourier (delay) modes. 

The optimal quadratic estimator formalism requires a data weighting matrix $\mathbf{R}$ and power spectrum normalisation matrix $\mathbf{M}$ to be specified. In the optimal case, we would weight the data by the inverse covariance, $\mathbf{R} = \mathbf{C}^{-1}$. The true covariance of the data is hard to model for 21cm experiments however, and using empirical estimates can induce disastrous signal loss in the OQE \citep{Ali:2018erratum}. Instead, we choose a sub-optimal but conservative (and lossless) set of identity weights, $\mathbf{R} = \mathbf{I}$. The choice of normalisation matrix determines the window function for each bandpower, and can be used to trade off the size of the errorbars against the degree of correlation between the bandpowers. We choose an intermediate case of $\mathbf{M} = \mathbf{I}$. More information on OQE and the notation used here can be found in \cite{2015ApJ...809...61A}.

\begin{figure*}
\begin{center}
\includegraphics[width=1.7\columnwidth]{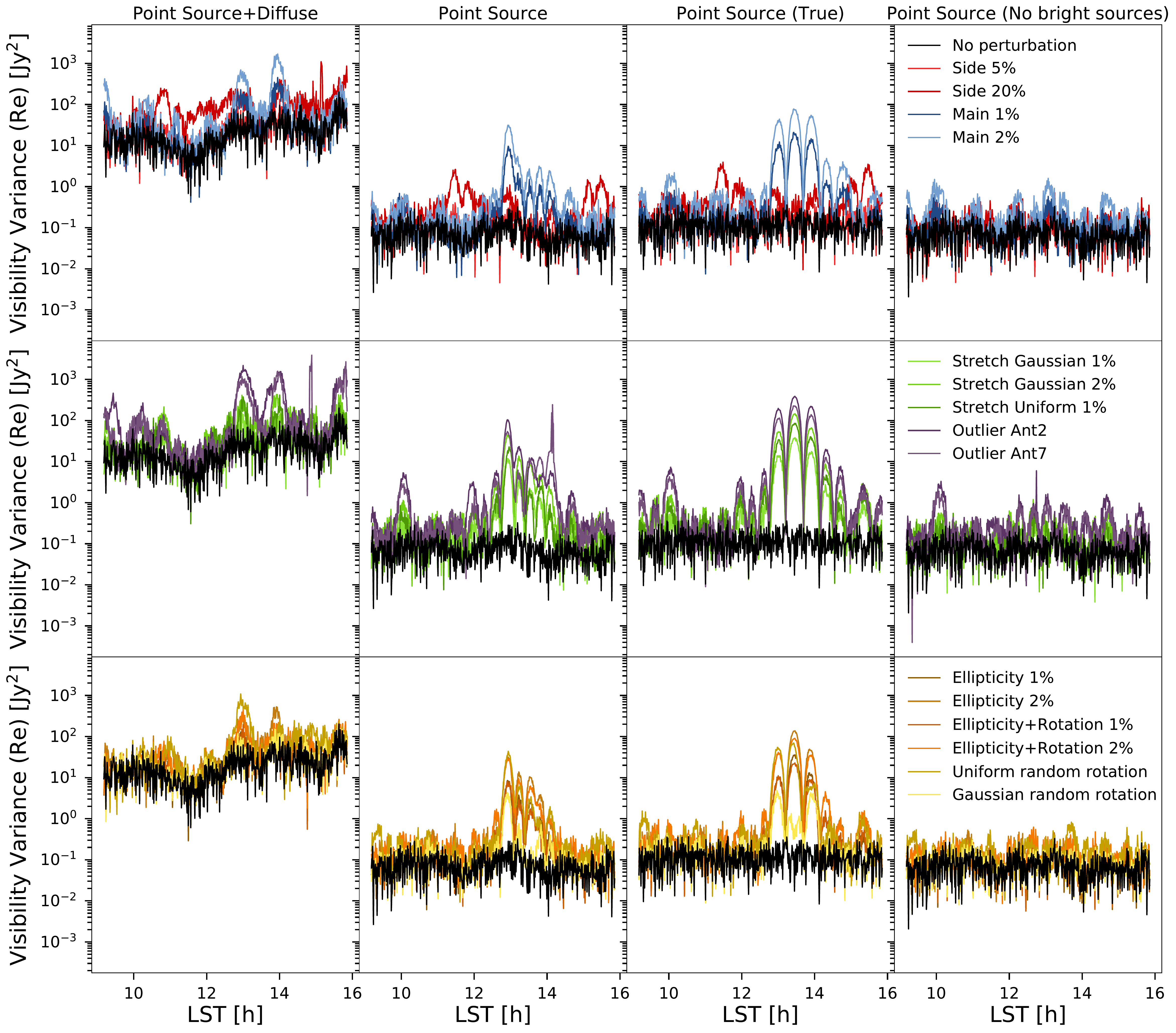}
\caption{The variance of the real part of the calibrated (columns 1, 2, and 4) and true (column 3) visibilities for a redundant group with baseline length $14.6$m and angle $0^{\circ}$ (from the E-W direction), containing 7 baselines. The results for different types of non-redundancy are shown in each row, and for different combinations of foregrounds in each column. The third column shows the variance for the true (simulated) visibilities, without any gain factors or calibration applied; it is a measure of the intrinsic variance in visibilities caused by the primary beam variations. Also note the important effect of bright sources on the variance -- when Centaurus A (flux 1937 Jy at 100 MHz) passes through the mainlobe at LSTs between $13-14$ hr, the variance increases by two orders of magnitude in the point source only case for most types of non-redundancy (second column).}
\label{fig:visvariance}
\end{center}
\end{figure*}

We form power spectra between all pairs of baselines within each redundant group, including for each baseline with itself. As noted above, in order to avoid a noise bias, each baseline pair is actually formed from neighbouring time samples, which have independent noise but essentially the same sky signal ($\Delta t = 40$ sec, compared with a primary beam crossing time of $\sim 40$ min). Both incoherent and coherent power spectra are calculated for each redundant group.

\vspace{-1em}
\section{Results}
\label{sec:results}

In this section, we examine the effects of each type of primary beam non-redundancy on the calibrated visibilities (Sect.~\ref{sec:redgrpvariance}), the gain solutions (Sect.~\ref{sec:gains}), the delay spectra (Sect.~\ref{sec:decorr}), and finally the recovered EoR power spectrum (Sect.~\ref{sec:eor}). {We also consider the effects of additional frequency-dependent sidelobes in Sect.~\ref{sec:freqsl}, and the smoothing of the gain solutions in Sect.~\ref{sec:gainsmooth}, as well as the effect of a larger array size (Sect.~\ref{sec:bigarray}).}

\begin{figure*}
\begin{center}
\includegraphics[width=1.78\columnwidth]{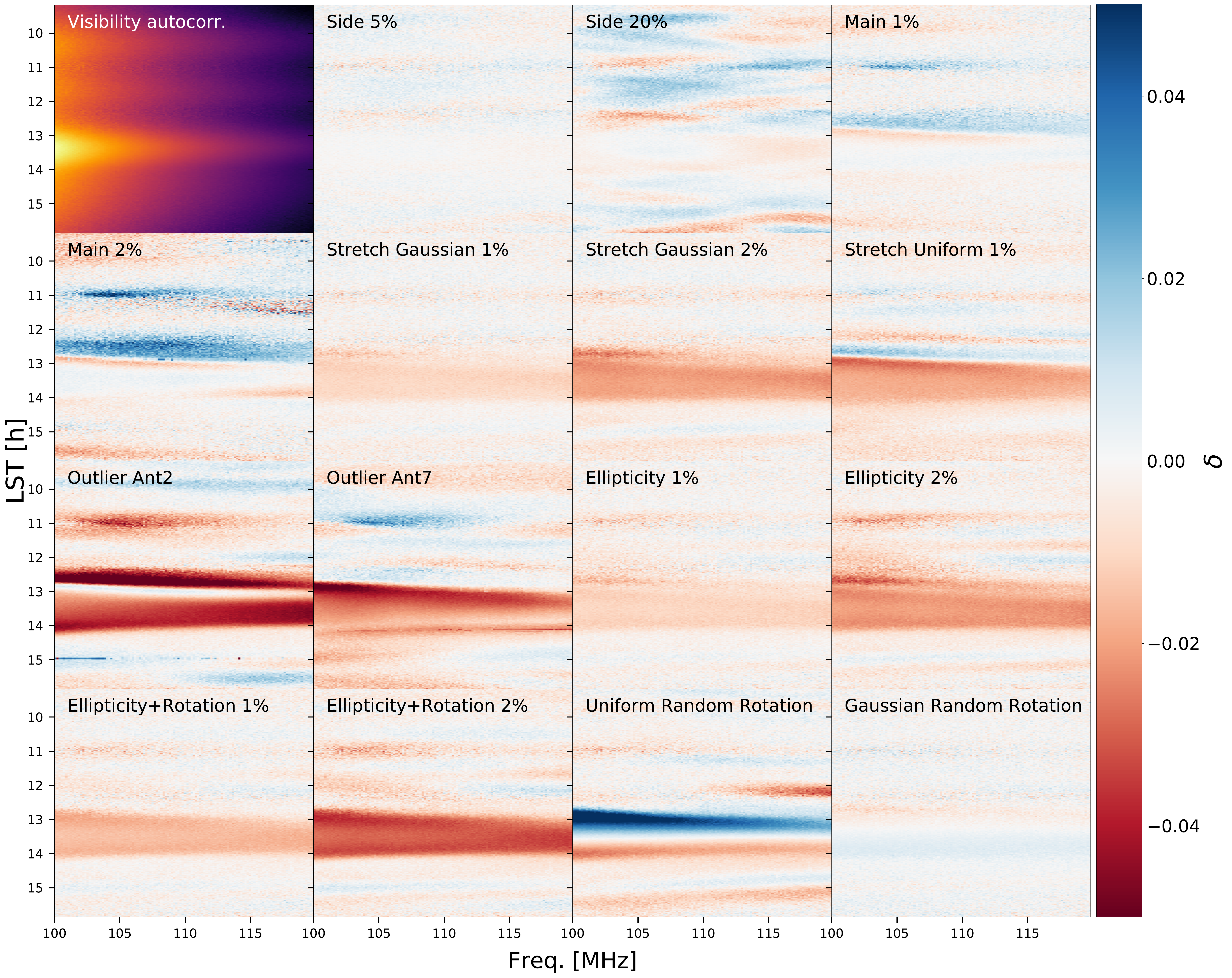}
\includegraphics[width=1.76\columnwidth]{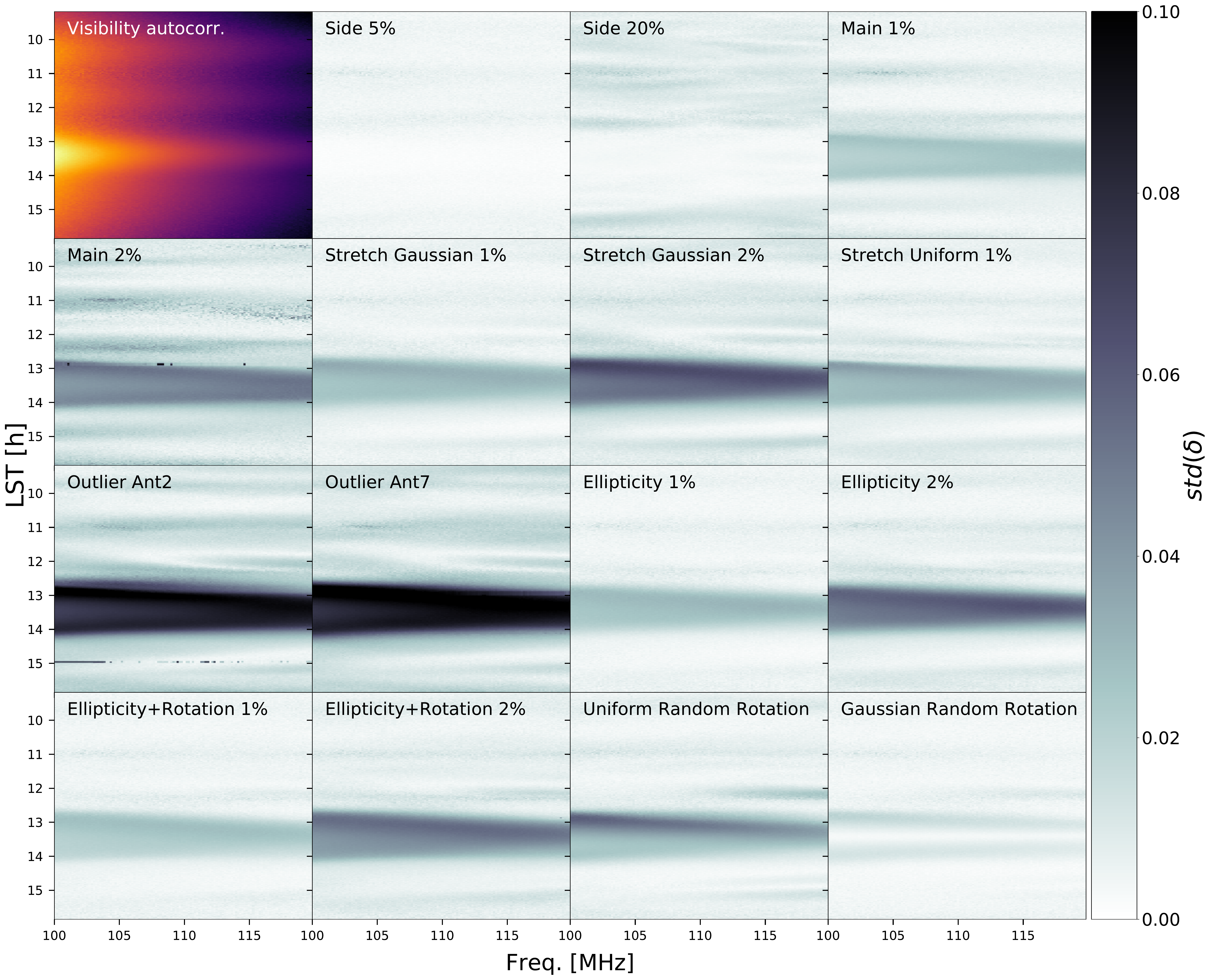}
\caption{{\it (Upper panel):} Fractional deviation of gain solutions after redundant calibration with respect to the true (input) gains, ($\delta_i=(g^{\rm cal}_i / g_i) - 1$), for antenna $i=1$ in the point source-only case. The different panels are for different types of non-redundancy except the top left panel, which shows the visibility autocorrelation on a different colour scale. {\it (Lower panel):} Same as above but now showing the standard deviation of $\delta$ over all 10 antennas. Note the glitches at $\sim 13$ h in the Main 2\% and Outlier Ant2 panels; these are caused by outlier gain solutions that were not detected by the filter (see Sect.~\ref{sec:redcal}).}
\label{fig:gaindiff}
\end{center}
\end{figure*}

\subsection{Variance of visibilities within a redundant group}
\label{sec:redgrpvariance}


In this section we investigate the size and structure of the additional intra-baseline variance introduced by different kinds of primary beam non-redundancy.

Following calibration, the visibilities within a redundant baseline group are typically also averaged together, either before or after forming power spectra, to improve the signal-to-noise on the final power spectrum. A perfectly calibrated, perfectly redundant array would be expected to have visibilities within a redundant group that differ only due to instrumental noise, which integrates down rapidly as more redundant baselines are included in the average. One of the effects of non-redundancy is to introduce additional visibility variations between baselines however. These are in part due to gain errors introduced by the redundant calibration, and partially due to the intrinsic differences between baselines due to the differing primary beams. At a minimum, this is expected to contribute additional variance in the averaged power spectrum, over and above the instrumental noise. Redundant calibration methods necessarily combine information from all baselines within each redundant group however, and so some level of correlation in these variations is also to be expected. This can potentially introduce spurious additional structure in the averaged power spectrum.

Fig.~\ref{fig:visvariance} shows the variance as a function of time for the calibrated visibilities in a single redundant baseline group, calculated across all baselines in the group, but for a fixed frequency of $110$ MHz. The upper row of this figure shows different levels of sidelobe and mainlobe perturbations, the middle row is for beams with a stretch factor applied, and the lower row is for elliptical/rotated primary beams. The columns shows different combinations of foregrounds that were included in the simulation: diffuse Galactic emission + point sources (first column); point sources only (second and third columns); and points sources only, but with the 20 brightest sources removed (fourth column). Each case shows the variance of the visibilities following redundant calibration and degeneracy fixing, apart from the third column, which shows the variance for the true (simulated) visibilities, without any gain factors or calibration applied. This is a measure of the {\it intrinsic} variance in visibilities caused by the primary beam variations only, separate from any calibration errors. The results for a perfectly redundant simulation are shown in black in each panel.

First, we note the structure of the variance in the perfectly-redundant `no perturbation' case. This is noise-like, as expected, and variations as a function of LST are caused only by the variations in sky brightness, which change $T_{\rm sys}$ and therefore the instrumental noise level. The noise level is significantly higher in the diffuse + point source case, as diffuse emission dominates the sky brightness.

In the case where there are no bright sources in the simulation (fourth column), we see that the variance is also mostly noise-like, with typical values on the order of $0.1$ Jy$^2$ for the whole range of LST and for all types of non-redundancy. There is some excess variance over the perfectly redundant case however, including some structure that corresponds to the brighter of the remaining point sources passing through the primary beam. These structures are most evident in the cases where there are outlier antennas, and also when the mainlobe perturbation is largest (the $2\%$ case). This suggests that the large numbers of fainter sources on the sky do not induce strong non-redundancy effects, although there is clearly still some additional variance that contributes at a level of a few times the instrumental noise level.

\begin{figure*}
\begin{center}
\includegraphics[width=1.65\columnwidth]{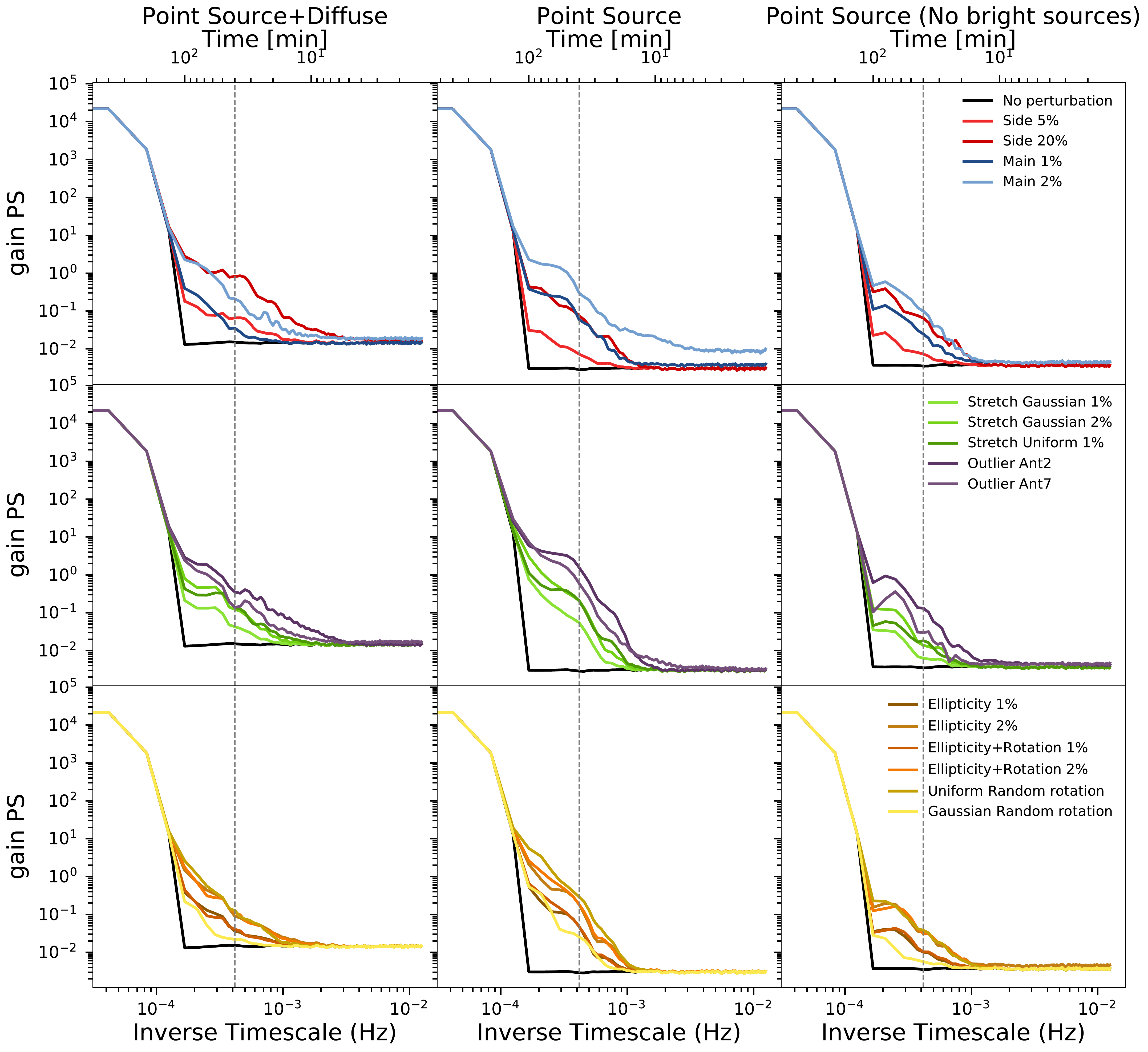}
\caption{The gain power spectrum as a function of LST. The panels represent the same {scenarios as in} Fig.~\ref{fig:visvariance}. Here, we average the gain power spectrum incoherently for all frequency and antennas. The vertical line show the 40 min time scale which is roughly the beam crossing time. The black solid line is for the perfectly redundant case. But, the non-redundancy generates the structures in temporal solution which manifest as a bump in the gain power spectrum. This temporal correlation increases when we increase the level of non-redundancy in the simulation. The noise floor in the gain power spectrum increases by an order of magnitude when we add the diffuse emission in the simulations}
\label{fig:gainps}
\end{center}
\end{figure*}

The second column of Fig.~\ref{fig:visvariance} shows the results after adding the brightest sources back into the simulation. We see a substantial increase in the variance for most types of non-redundancy, with an almost two orders of magnitude change in the LST range $13-14$ h for all cases except the sidelobe  perturbations. This is caused by the very bright source Centaurus A (Cen~A; flux 1937 Jy at 150 MHz) transiting through the mainlobe in this LST range, as illustrated in Figs.~\ref{fig:tracks} and \ref{fig:pbflux}. The brightness of this source means that it contributes a large fraction of the total flux for each visibility, dominating over other contributions such as fainter point sources or noise, and therefore highlighting any differences in the primary beam patterns between antennas at specific (localised) zenith angles as a function of LST. As such, bright sources can be used specifically to map the primary beams \citep{2012AJ....143...53P, 2020ApJ...897....5N}, and so it is understandable that the additional variance due to primary beam non-redundancy should be maximised when such a source transits.

\begin{figure*}
\begin{center}
\includegraphics[width=1.8\columnwidth]{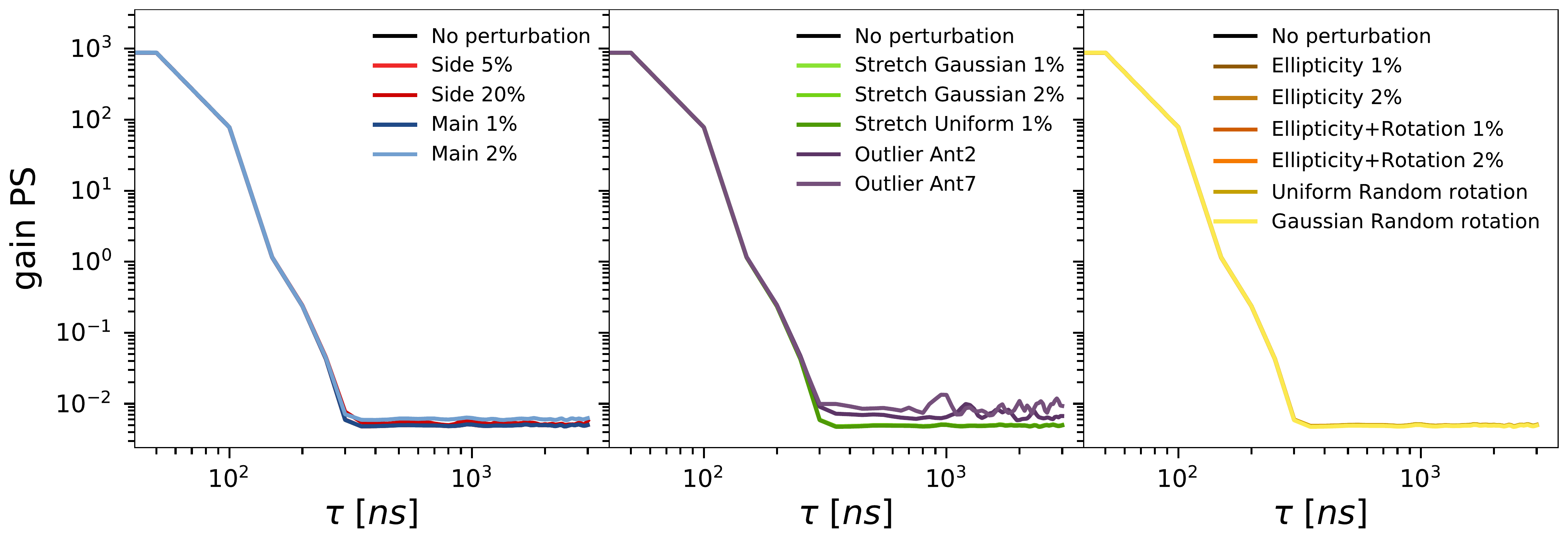}
\caption{{The gain power spectrum as a function of delay, $\tau$, in this case for only the Point Source + Diffuse scenario, following incoherent averaging of the power spectra for all LSTs and antennas. In comparison with the temporal power spectrum shown in Fig.~\ref{fig:gainps}, there is little difference in structure between the different non-redundant cases.}}
\label{fig:delaypsgain}
\end{center}
\end{figure*}

Most of the different types of non-redundancy result in a repeated, oscillating structure in the variance around the time of the Cen~A transit, corresponding to when the source passes through the sidelobes, nulls, and then through the mainlobe and out the other side. The exception is in the sidelobe perturbation cases, where the variance is only enhanced in the LST ranges around $11-12$ h and $15-16$ h. Effects of other bright sources can also be seen at a lower level, such as the feature in the variance for the Outlier cases around 10 h, presumably related to the presence of sources A and B in the mainlobe.

For comparison, the third column of Fig.~\ref{fig:visvariance} shows the intrinsic variance of the true (simulated) visibilities in the point source-only case. Here, we see that the variance is enhanced by almost two orders of magnitude compared with the calibrated data at LSTs of around $13-14$ h, where bright sources Cen~A and source C are passing close to/through the mainlobe, but practically no enhancement at all other LSTs. This is true for most types of non-redundancy, and suggests that the redundant calibration procedure is able to absorb some of the effects of primary beam non-redundancy -- but only when the visibility model is dominated by a single bright source. In fact, it can be seen from Eqs.~\ref{eq:gains} and \ref{eq2} that variations between the primary beams can be absorbed {\it exactly} into the gains in the special case where the sky model, $V_{i j}^{\rm true}$, contains only a single point source. A fictitious perfectly-redundant visibility model can therefore be obtained for all redundant groups in this case, assuming no other sources of non-redundancy are present. We also inspected the intrinsic variance for the diffuse + point source simulations, finding a similar behaviour when the brightest point sources are transiting, albeit with a smaller level of enhancement, and with no enhancement in the variance otherwise.

Finally, the left column of Fig.~\ref{fig:visvariance} shows the variance after adding diffuse emission. We see that the overall noise level increases by around two orders of magnitude. However, the features due to the strong source Cen~A, and the fainter sources A and B, are still present in those LST ranges, albeit not quite as prominently. An additional trend with LST is also observed as the Milky Way rises, with a substantial increase in variance seen from 15.5 h onwards as the plane of the Galaxy approaches the mainlobe (it would cross the mainlobe at an LST of around $17-18$ h).

\begin{figure*}
\begin{center}
\includegraphics[width=1.95\columnwidth]{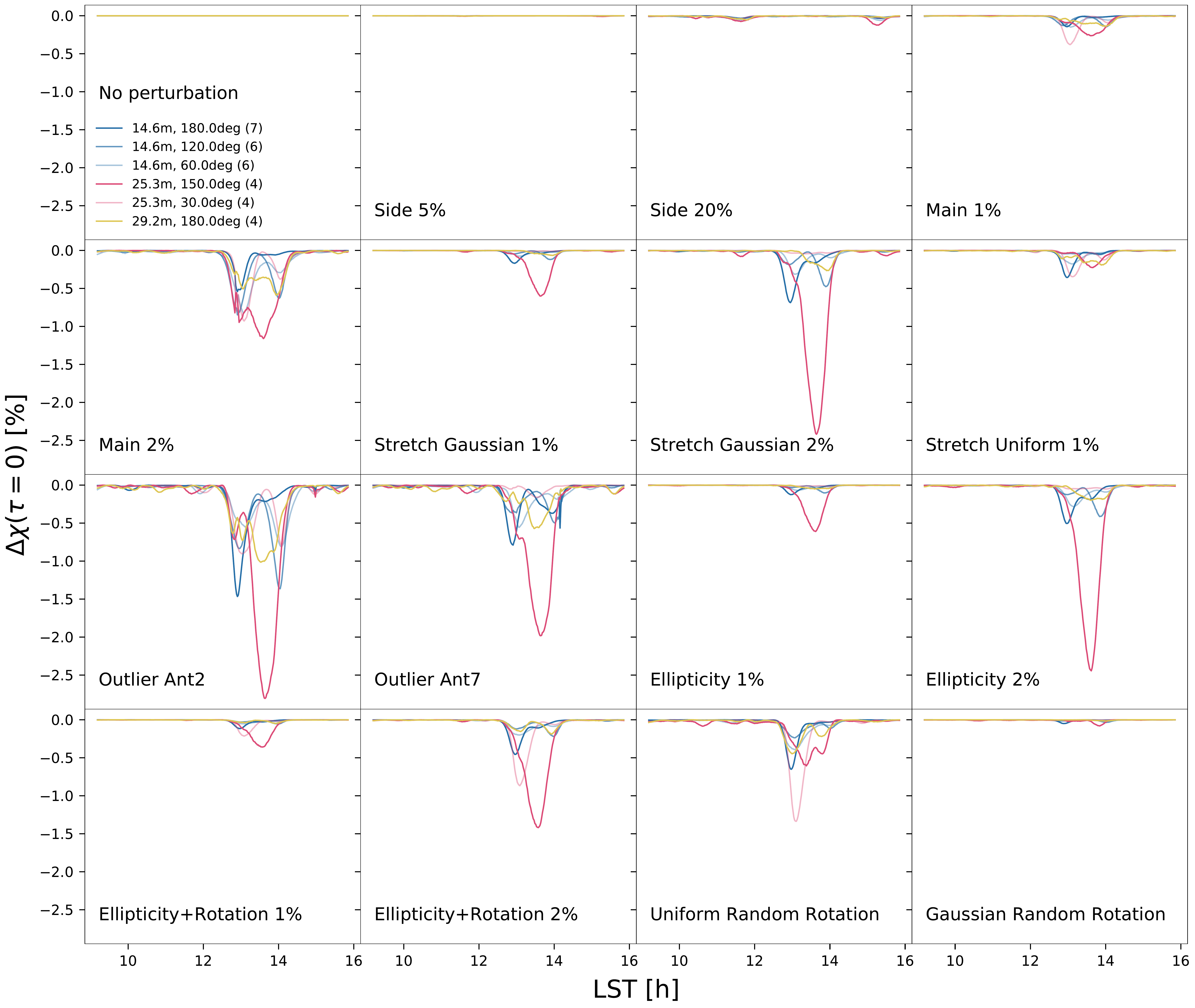}
\caption{Fractional power spectrum decoherence statistic $\Delta \chi(\tau=0, t)$, shown as a percentage, for the different types of non-redundancy in the point source-only case. Different redundant groups are shown with different colors; only redundant groups with at least 10 baselines are included. In all cases, the most substantial deviations from perfect correlation between the coherent and incoherent power spectra appears in the LST range $\sim 13-14$ h, where the brightest source in the data, Cen~A, transits the mainlobe.}
\label{fig:decorr1}
\end{center}
\end{figure*}

\begin{figure*}
\begin{center}
\includegraphics[width=1.95\columnwidth]{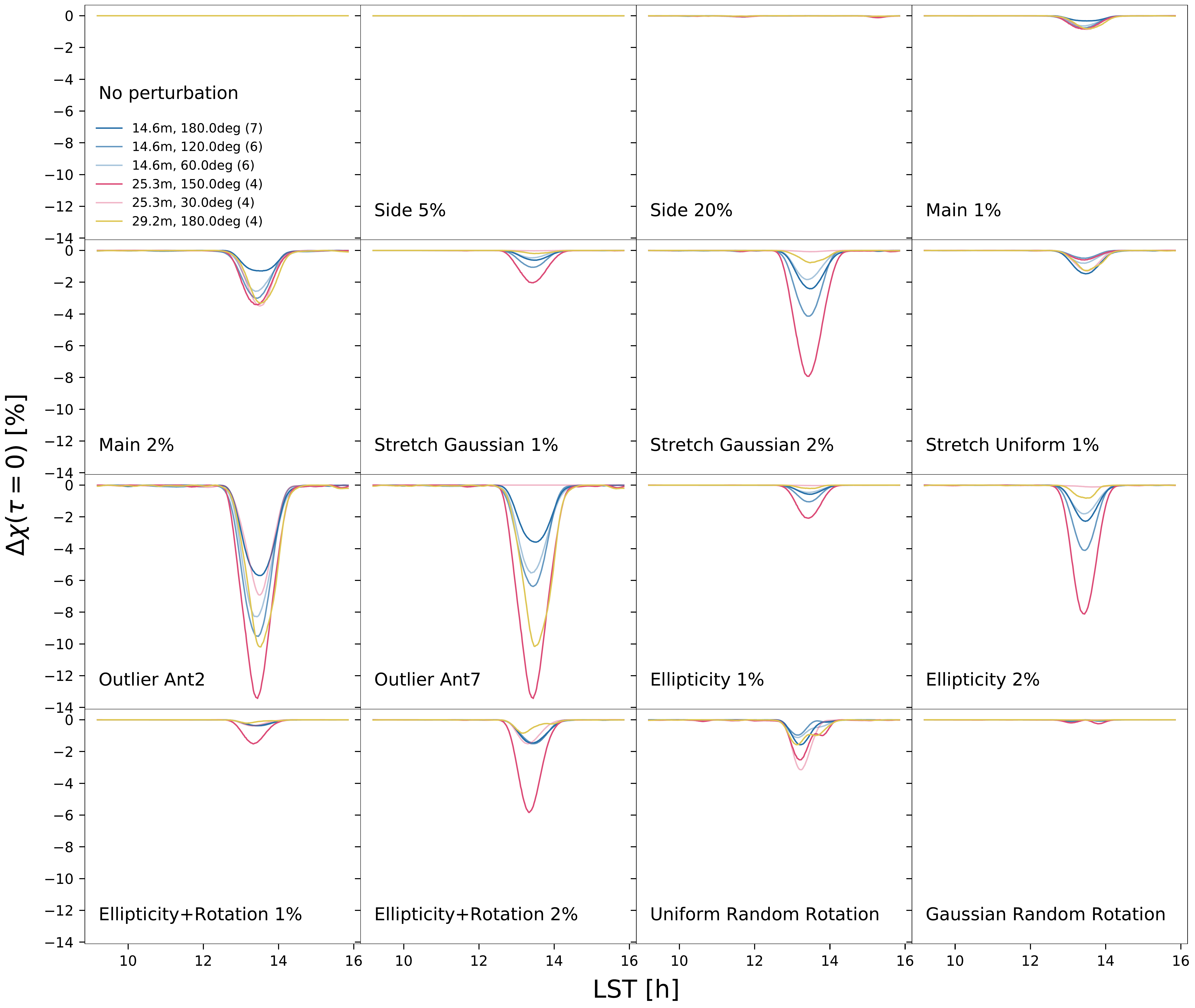}
\caption{Same as Fig.~\ref{fig:decorr1}, but for power spectra calculated from the true (simulated) visibilities, without any gain factors or redundant calibration applied to them. This is a measure of the intrinsic decoherence caused by the variations in the primary beam between antennas. Note the substantially increased range on the $y$ axis.}
\label{fig:decorr1a}
\end{center}
\end{figure*}

\begin{figure*}
\begin{center}
\includegraphics[width=1.95\columnwidth]{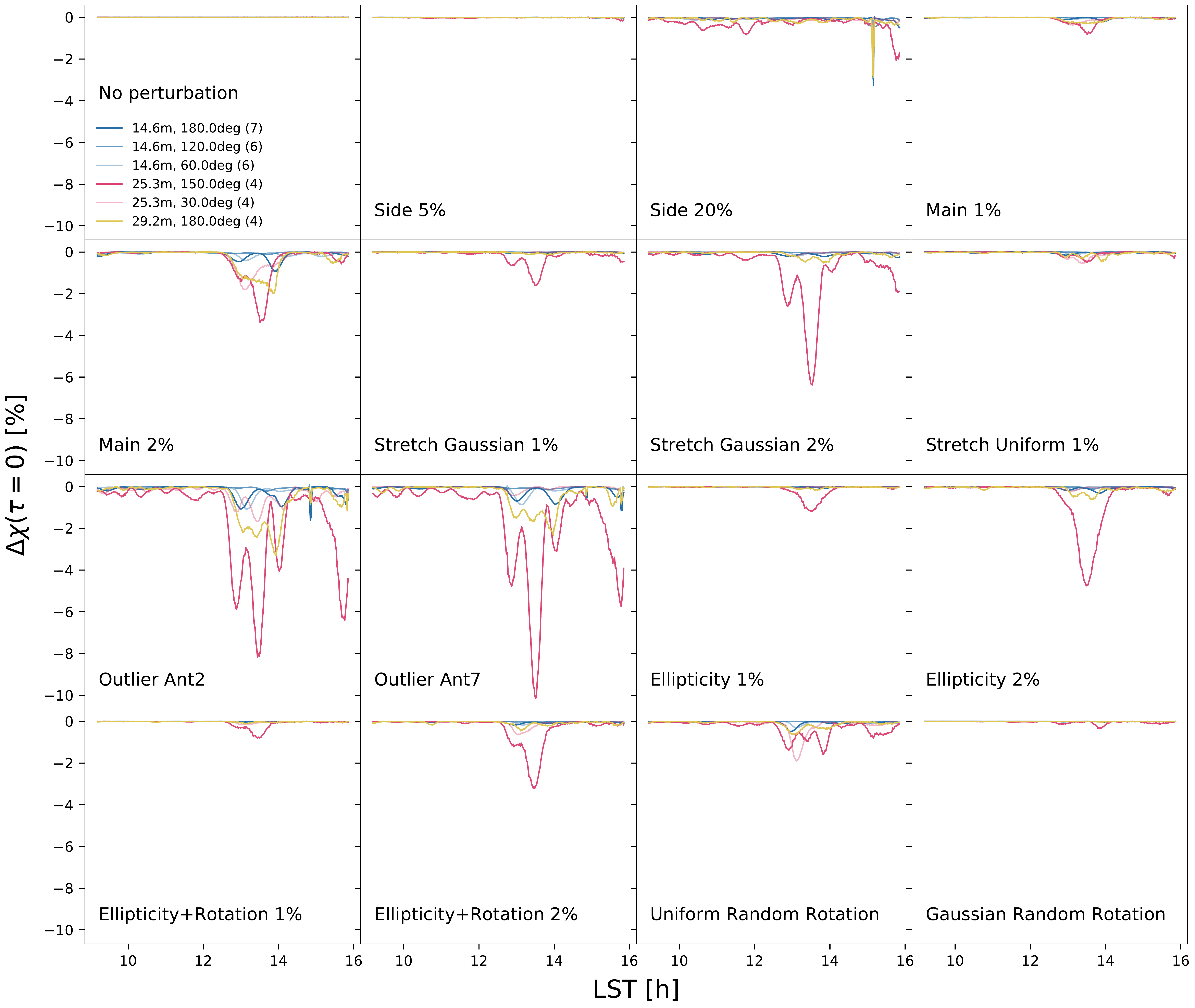}
\caption{Same as Fig.~\ref{fig:decorr1a}, but now with diffuse emission and point sources included in the simulations.}
\label{fig:decorr2}
\end{center}
\end{figure*}
\subsection{Temporal and spectral structure of the gain solutions}
\label{sec:gains}

In this section, we study the temporal and spectral variations of the antenna gain solutions, $\{g^{\rm cal}_i\}$, output by the redundant calibration process. It is first useful to define the fractional gain error as the fractional deviation from the true (input) gain, $\delta_i = (g^{\rm cal}_i/g_i) - 1$, where $g_i$ is the true input gain, and $i$ labels the antenna.

Fig.~\ref{fig:gaindiff} shows $\delta$ as a function of LST and frequency for antenna 1, which is located at the centre of the array. Each panel of Fig.~\ref{fig:gaindiff} shows $\delta$ for one of the 15 different types of non-redundancy considered in this study, except the top left panel which shows the amplitude of the visibility autocorrelation for the antenna, $V_{11}$, shown on a different colour scale, to give some idea of the total sky brightness as a function of LST and frequency.

We see that $\delta$ largely varies between $-5\%$ to $+5\%$, with only a few scenarios briefly saturating the colour scale when Cen~A transits through the mainlobe at around $13-14$ h. In the majority of cases, the maximum deviations are associated with the Cen~A transit, with significant variations in the behaviour of $\delta$ from case to case. For example, the Uniform Random Rotation case shows $\delta$ cross from positive to negative during the transit, while the Ellipticity + Rotation 2\% case shows a consistent negative signature. 

In comparison, the Side (sidelobe) cases show considerably more structure as a function of LST, and in fact exhibit the lowest level of $\delta$ during the Cen~A transit. This is to be expected as Cen~A leaves the sidelobes (which are non-redundant) and enters the mainlobe (which is almost perfectly redundant in this case). Judging by this behaviour, it seems that much of the gain error at other times must be caused by Cen~A in this case, as $\delta$ {\it only} reduces when the source is safely inside the mainlobe.

There is also a notable difference in the pattern of the gain errors depending on which distribution the primary beam non-redundancies are drawn from. There are significant differences between the Stretch Gaussian 1\% and Stretch Uniform 1\% cases for example, while the more coherent Gaussian Random Rotation case is significantly more redundant than the Uniform Random Rotation Case.

To summarise the gain errors across all antennas, we also show the standard deviation of $\delta$ over all $10$ antennas in Fig.~\ref{fig:gaindiff} (lower panel), defined as $\sigma_\delta = \sqrt{{\rm Var}(\delta)}$. It can be seen that the structure of this plot is very similar to Fig.~\ref{fig:gaindiff}, and so our conclusions from above continue to hold. The values of $\sigma_\delta$ largely vary between $0$ to $0.1$, corresponding to a maximum 10\% gain error, with only a couple of cases saturating the colour scale during the Cen~A transit.

To further analyse the temporal variation of the gain solutions, in Fig.~\ref{fig:gainps} we show the temporal power spectrum of the redundant gain solutions output by {\tt redcal} (and, as above, following degeneracy fixing and outlier removal). The temporal gain power spectrum is calculated by first taking the Fourier transform of the gain solution in the LST direction for a particular antenna and frequency channel after multiplying by a Blackman-Harris taper, and then squaring the result. Here, we average the temporal gain power spectrum incoherently for all frequency channels and antennas. Different panels in Fig.~\ref{fig:gainps} are for different types of non-redundancy and foreground simulations, as mentioned in Sect.~\ref{sec:redgrpvariance}. The vertical black dashed line shows the approximate time scale ($\sim 40$ min) for a source to cross the mainlobe of the primary beam, while the black solid line in each panel shows the gain power spectrum for the perfectly redundant case.

In all cases, we see that there is essentially identical temporal structure above 100 minutes. This is caused by the modulation of the mean of the gain solutions ($\simeq 1$) by the taper, and is present even in the perfectly redundant case. For perfect redundancy, there is essentially no structure on shorter timescales however, with only a flat thermal noise floor present. The floor rises when diffuse emission is included due to the corresponding increase in $T_{\rm sys}$.

In the presence of primary beam non-redundancy, the gain solutions exhibit correlated structure on shorter timescales, typically of order the beam crossing time. In our simulations, the true gains are constant over time, so any structure present in the final gain solutions (beyond the noise and the taper effect mentioned above) is due to the effect of primary beam non-redundancy. 

The amount of additional structure in the {temporal} gain power spectrum increases when we increase the level of non-redundancy, as expected. The behaviour is quite similar for all types of non-redundancy and the three different foreground models considered here, with the main difference being the increased noise floor when diffuse emission is added, although an enhancement in $\sim\!100$ min. timescale structure (just before the taper feature) can be seen in going from the point source only case with no bright sources vs. the one with bright sources. Otherwise, the structure of the gain power spectrum between cases seemingly only differs in detail. (N.B. The Main 2\% case in the middle panel has an increased noise floor due to an unaddressed outlier gain solution.)

{Finally, in Fig.~\ref{fig:delaypsgain} we show the spectral power spectrum of the gain solutions as a function of delay, incoherently averaged over LST and antenna. In contrast to Fig.~\ref{fig:gainps}, there is little difference in the spectral gain power spectrum between the different types of non-redundancy. The main low-delay structure in the gains is almost identical between the different cases, with only small differences observed in the noise `floor' at higher delay for the two outlier antenna cases.}


\subsection{Decoherence of the delay power spectrum}
\label{sec:decorr}
Finally, we compare the coherently- and incoherently-averaged power spectra within each redundant group as a way of studying possible signal loss due to gain errors induced by the primary beam non-redundancies. Figs.~\ref{fig:decorr1} and \ref{fig:decorr2} show the results for the point source-only and point sources + diffuse emission simulations respectively. Fig.~\ref{fig:decorr1a} shows the same point source-only simulation but now using delay spectra calculated from the true (intrinsic) visibilities, with no gains or calibration applied. We plot the fractional difference between the coherently- and incoherently-averaged delay spectra,
\be
\Delta \chi(\tau, t) = \frac{P_{\rm co}(\tau, t) - P_{\rm inco}(\tau, t)}{\langle P_{\rm inco} \rangle_t} \label{eq:decorr}
\ee
at delay $\tau=0$ and LST $t$, where $P_{\rm co}$ and $P_{\rm inco}$ are the coherently- and incoherently-averaged delay spectra within a particular redundant group (Eqs.~\ref{eq:Pinco} and \ref{eq:Pco}), and $\langle P_{\rm inco} \rangle_t$ is the power spectrum averaged incoherently over the entire LST range of the simulated observations as well. Recall that $P_{\rm inco}$ is formed from the auto-baseline spectra only.

The rationale for studying this statistic is as follows (HERA Collaboration 2021, {\it in prep.}). In principle, the coherent average should provide a measurement of the power spectrum with improved signal-to-noise, as the noise from each visibility will coherently average down. Gain errors induce both phase and amplitude errors in the calibrated visibilities however. The amplitude errors effectively act as an additional source of noise, and are also expected to average down as more baselines are included in the coherent average, albeit slower than for thermal noise because the amplitude errors are correlated. The phase errors on the other hand can destructively interfere, effectively causing signal loss by decohering the signal.

By comparing the coherently-averaged power spectrum to the incoherently-averaged one (which does not suffer from this decoherence effect), we can get a handle on the magnitude of any signal loss in the power spectrum. It is particularly instructive to do this for the $\tau=0$ mode, as this is where most of the foreground power resides, and it therefore has a very high SNR. Any signal loss would be most noticeable for this mode. Plotting the decoherence statistic $\Delta \chi$ as a function of LST allows us to see how any signal loss varies as different foreground structures pass through the beams. The denominator in Eq.~\ref{eq:decorr} is chosen as a time average over the whole observation range to provide a stable baseline level as a reference (and to avoid zero crossings), but it should be noted that the quantitative results depend on the length of the time averaging domain ({\it S.~Singh, priv. comm.}).

Turning to Fig.~\ref{fig:decorr1} (for the point sources-only sky model), it can be seen that practically all non-redundant cases exhibit appreciable decoherence at $\sim 13-14$ h, when Cen~A is transiting the mainlobe. The size (and shape) of the effect differs between baselines and different types of non-redundancy, with the longer baselines exhibiting values of $\Delta \chi$ as large as $-2.5\%$. There appears to be a significant orientation effect, which can be seen by comparing the 14.6m baseline groups with three different orientations -- the $180^\circ$ group (perfectly E-W aligned) tends to produce a larger (more negative) value of $\Delta\chi$ that peaks at earlier LST, while the $120^\circ$ group reliably shows a less negative $\Delta\chi$ that peaks later. A similar orientation-dependent effect is seen in other groups, with the 25.3m $150^\circ$ group almost always exhibiting a significantly more negative $\Delta\chi$ than the $30^\circ$ group. Interestingly, this is reversed in a couple of cases, e.g. Uniform Random Rotation.

Another notable feature of Fig.~\ref{fig:decorr1} is that in some cases, all baseline lengths and orientations are affected by decoherence during the Cen~A transit (e.g. Mainlobe 2\%, Outlier Ant. 2), whereas in others (e.g Ellipticity 2\%) certain orientations are more or less unaffected. This suggests the possibility of performing a rough identification of which forms of primary bean non-redundancy are likely to be in operation by comparing the LST-dependence of the decoherence statistic between redundant groups as a bright source transits. Also notable is the very low level of decoherence in the Sidelobe perturbation cases, as well as the reduced level of decoherence in the Gaussian Random Rotation case compared with the Uniform Random Rotation case.

Finally, it can be seen that only the brightest source, Cen~A, appears to generate any appreciable decoherence. Even the bright source A (see Figs.~\ref{fig:tracks} and \ref{fig:pbflux}), which peaks at around 9 h, induces little decoherence, even in the most extreme cases.

Fig.~\ref{fig:decorr1a} shows the decoherence statistic for the true visibilities, without any gain factors or redundant calibration applied to them. This gives us a handle on the intrinsic decoherence caused by the different types of primary beam non-redundancy, before any correcting factors or complications due to the gain solutions are taken into account. We see that the features appear in the same LST range as Fig.~\ref{fig:decorr1}, around 13-14 h, where the bright source Cen~A is close to the mainlobe. They are smoother and more symmetric however. The maximum deviation is also significantly larger, reaching a level of decoherence as great as $-14\%$ in the Outlier cases, compared with $-2.5\%$ for the same data when gains are included and the redundant calibration solutions are applied. As discussed in Sect.~\ref{sec:redgrpvariance} and shown in Fig.~\ref{fig:visvariance}, the intrinsic variance due to the primary beam non-redundancies can be quite large when the brightest sources are passing close to the mainlobe. There is a substantial reduction in the size of this effect when the redundant gain calibration is applied however, since the effect of the variations between the primary beam values at the location of the dominant point source can be partially absorbed by the gain solutions, effectively making the calibrated visibilities more redundant than they really are.

A similar behaviour is also seen in real HERA data, for example in Fig.~10 of \cite{2020arXiv200308399D}, where the $\chi^2$ values of the gain solutions are markedly lower when the bright source Fornax~A is moving through the mainlobe -- suggesting greater redundancy of the solutions. A double-peaked structure is also observed when Fornax~A passes through the sidelobes, which we also see in most of the cases in Fig.~\ref{fig:decorr1} after the redundant calibration solutions have been applied. In comparison, in the same LST range we see only a single, symmetric peak for the intrinsic decoherence (see Fig.~\ref{fig:decorr1a}), where no gains or calibration have been applied. This suggests that the redundant calibration is able to find solutions that are artificially more redundant (i.e. more redundant than the intrinsic visibilities) when a bright source is in the mainlobe, but less so when it is in the sidelobes.

The behavior of the decoherence in presence of diffuse emission as well as point sources (now for the calibrated data again) is shown in Fig.~\ref{fig:decorr2}. This is qualitatively quite similar to the results in the point source-only case, but the detailed shapes of the curves are different, with more fluctuations, and the minima are much deeper ($\Delta \chi \sim -10\%$) in the most extreme cases. There is generally more low-level structure across the LST range, particularly for the two Outlier cases, and to some degree for the Stretch Gaussian cases too. Similar observations hold about the orientation-dependence of the decoherence statistic as in Fig.~\ref{fig:decorr1}. Additionally, several of the cases now exhibit additional structure in $\Delta\chi$ at the top end of the LST range, as the Galactic plane begins to rise.

\subsection{Frequency-dependent sidelobe perturbations}
\label{sec:freqsl}

{Next, we study the effect of an explicitly frequency-dependent sidelobe perturbation of the form given in Eq.~\ref{eq:slfreq}, which is listed as Case 1(b) in Table~\ref{table1}. The red and blue lines in Fig.~\ref{fig:decorfreqsl} show the decoherence statistic $\Delta\chi(\tau=0)$ with and without the additional frequency dependence of the sidelobe perturbation respectively. Illustrative waterfall plots of the primary beams for both cases was shown in the two rightmost panels of Fig.~\ref{fig:polyfreq}), albeit for a more extreme version of Case 1(b).

From Fig.~\ref{fig:decorfreqsl} (left panel), we can see that the amount of decorrelation increases slightly after adding the additional frequency dependence to the beam model. The curve for Case 1(b) is also somewhat noisier, and reaches a minimum at an earlier LST value around the bright source Cen~A. This is likely a result of additional (non-redundant) structure being introduced into the visibilities due to the increased complexity of the non-redundant primary beams, a difference that can also be observed in the variance of the visibilities within each redundant group (see Fig.~\ref{fig:decorfreqsl}, right panel). This additional structure causes a reduction in the effectiveness of the decoherence-suppressing effect of redundant calibration when a bright source is in the field of view, which was discussed in the previous section. Note that there is also a spurious positive spike in the value of $\Delta\chi$ at around 14.5 h, which is unphysical (due to an outlier).

\begin{figure*}
\begin{center}
\includegraphics[width=1.\columnwidth]{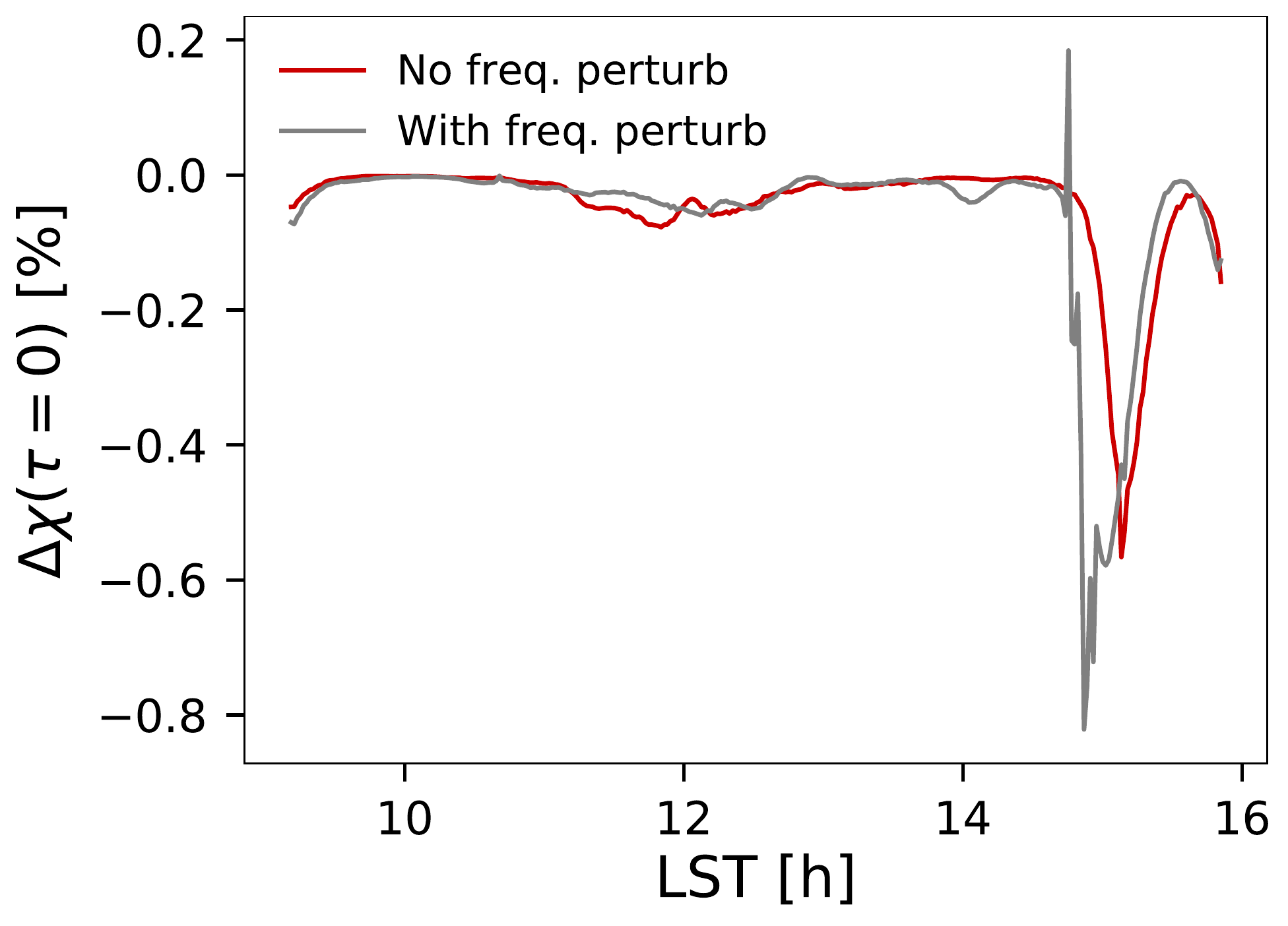}
\includegraphics[width=1.\columnwidth]{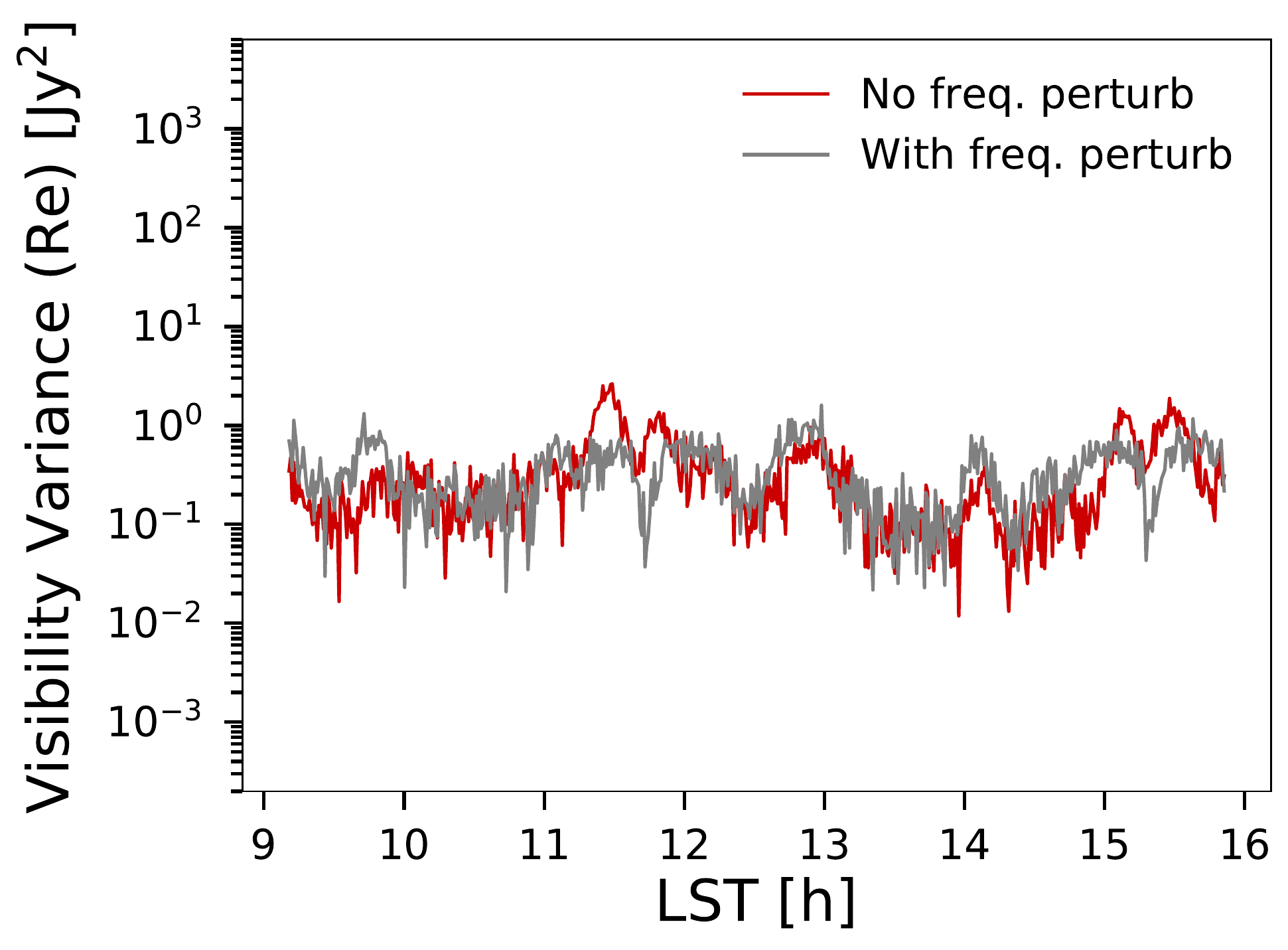}
\caption{{{\it (Left)}: Comparison of the decoherence statistic $\Delta\chi(\tau=0)$ for the standard sidelobe perturbation (Case 1a, $\sigma_{\rm SL} = 0.2$; red line), and the same case with an additional frequency dependence (Case 1b; grey line), for point source-only simulations. The statistic is shown for the 14.6m, $120^{\circ}$ redundant baseline group. Note the increased noisiness of the grey curve, which is also slightly deeper and has an earlier onset around the bright source Cen~A. {\it (Right)}: The variance of the visibilities for the same two cases, calculated as in Fig.~\ref{fig:visvariance}.} }
\label{fig:decorfreqsl}
\end{center}
\end{figure*}

Introducing a significant additional frequency dependence to the sidelobes has only modified the decoherence effect slightly in this case. Different types of added frequency dependence, e.g. that affect the mainlobe instead, may cause more significant differences, as might different functional forms for the frequency dependence. The increased noisiness of the decoherence statistic may also be mitigated if a larger redundant baseline group is averaged over. A thorough investigation of these questions is deferred to future work. We point out that this will need to find a constrained set of (ideally more realistic) frequency-dependent perturbations however, as otherwise the number of different combinations of possible zenith angle-, antenna-, and frequency-dependent non-redundancies will become unwieldy.}

\begin{figure}
\begin{center}
\includegraphics[width=1.\columnwidth]{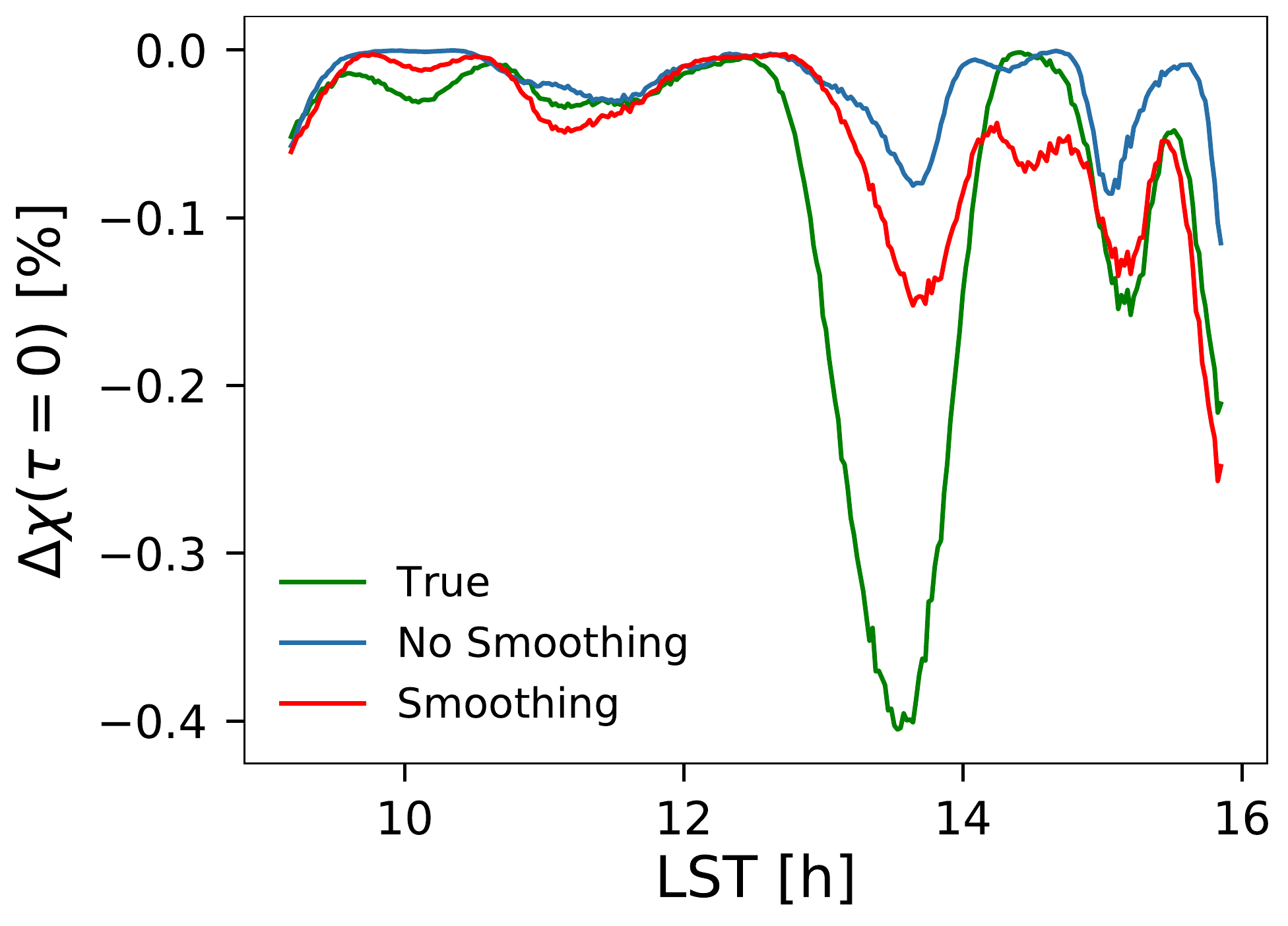}
\caption{{Comparison of the decoherence statistic $\Delta\chi(\tau=0)$ for unsmoothed gain solutions (blue), and gain solutions smoothed on a time scale of approximately 2.2 hours (red), for the 14.6m, $120^{\circ}$ redundant baseline group (6 baselines) in the Stretch Gaussian 2\% case, for point source+diffuse simulations. The green line shows the intrinsic decoherence that occurs when the visibilities are redundantly-averaged after the true gain solutions are applied. Gain smoothing brings the gain solutions closer to their true values, but results in increased decoherence.}}
\label{fig:gainsmooth}
\end{center}
\end{figure}

\begin{figure*}
\begin{center}
\includegraphics[width=1.78\columnwidth]{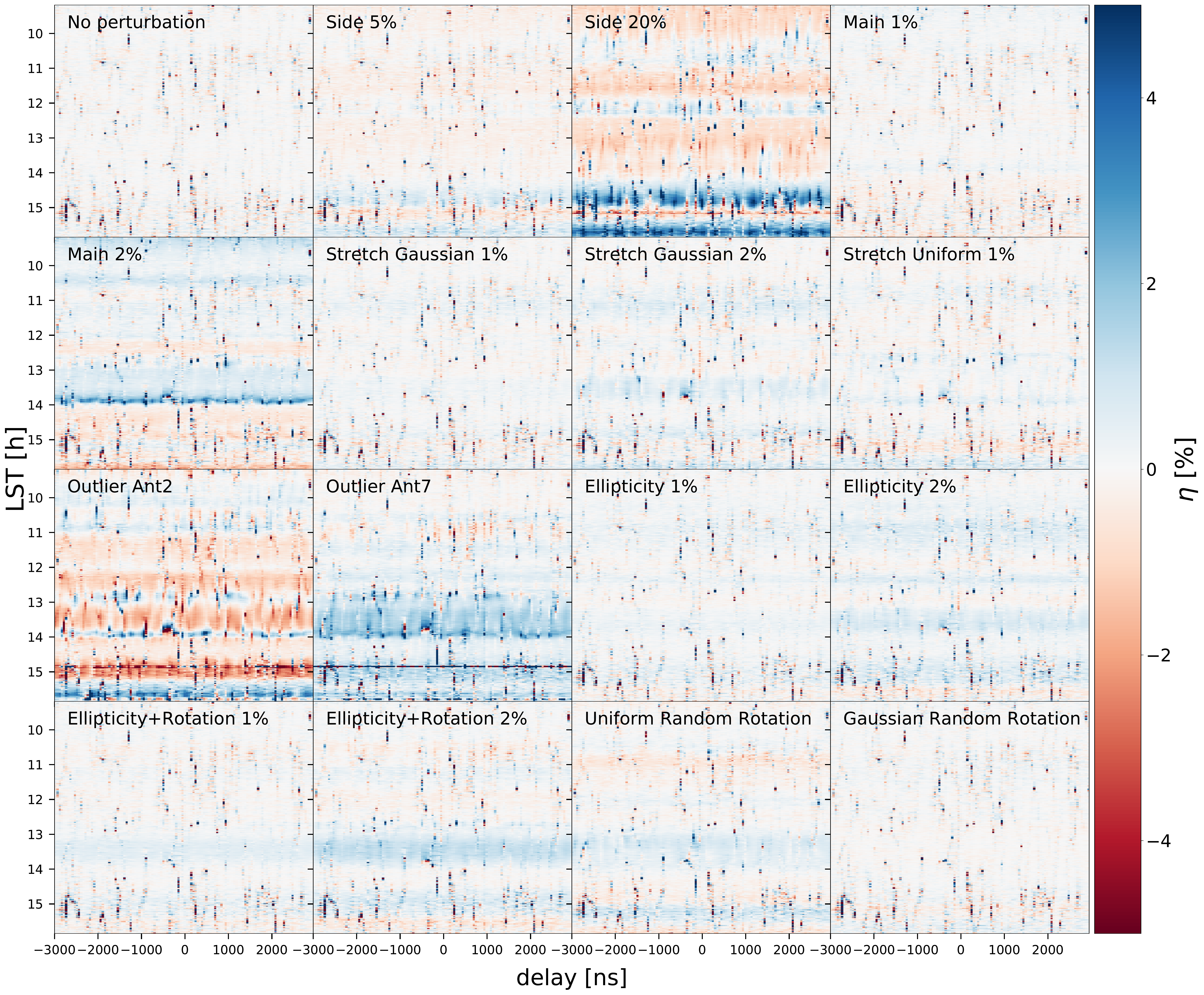}
\caption{Fractional deviation of the EoR power spectrum from its true value, $\eta=(P / P_{\rm true}) - 1$ (shown as a percentage), after redundant calibration solutions from the point source + diffuse simulations have been applied. This figure shows the results for the real part of the incoherently-averaged delay spectra for the 14.6m, $180^{\circ}$ redundant group. The different panels are for different types of non-redundancy. The localised spikes are mostly caused by zero-crossings in $P_{\rm true}$.}
\label{fig:eorpsdiff}
\end{center}
\end{figure*}


\subsection{Temporal smoothing of gain solutions}
\label{sec:gainsmooth}

{We now study the effect of smoothing the gains on the decoherence effect. Temporal gain smoothing is used to reduce spurious structure in the gain solutions produced by redundant calibration; apart from some short timescale variations due to a hardware cooling cycle, the HERA gains are expected to be essentially stationary over timescales of a few hours \citep{2020ApJ...890..122K}, and so removing structure with faster variations than this is expected to result in more accurate gain solutions. Given a typical beam-crossing time of $\sim 40$ mins, this should also reduce gain errors caused by non-redundancy as bright sources pass through the field of view.

To reproduce this treatment, we smooth the gain solutions from our pipeline over a timescale of 2.2 hours. We do not apply any smoothing in the frequency direction however. Fig.~\ref{fig:gainsmooth} shows the values of the decoherence statistic, $\Delta \chi$, after smoothing the antenna gains (red line) for the particular case of the Stretch Gaussian $2\%$ non-redundant primary beams. The blue line shows the corresponding result without gain smoothing, and the green line shows the `intrinsic' decoherence when the true gains are applied to the visibilities instead.

We see that after smoothing, $\Delta \chi$ tends towards the intrinsic (true) curve, which is due to the gain solutions becoming closer to their true values, as expected. Interestingly, this {\it increases} the level of decoherence however. This fits with our finding from Sect.~\ref{sec:decorr}, that redundant calibration erroneously absorbs some of the non-redundancy of the true visibilities; smoothing the gains reduces this effect, although it does not remove it altogether.}

\subsection{Modulation of the EoR power spectrum}
\label{sec:eor}

Gain errors caused by primary beam non-redundancy ultimately cause a modulation of the cosmological 21cm signal from Cosmic Dawn/EoR. To study this effect, we used {\it healvis} to generate a model EoR signal with a flat (white noise) power spectrum. The same array configuration, gains, and non-redundant primary beams were used as in the rest of this paper, but noise and foregrounds were not applied to the simulations. Instead, we took the calibration solutions that were derived from the simulations with noise and foregrounds (but no EoR) and applied them directly to the EoR simulations. We did not repeat the entire calibration process with all components included, as the EoR signal is small in comparison with the foregrounds and so should change the gain solutions only by a negligible amount.

After applying the gain solutions to the simulated EoR visibilities, we then calculated their power spectra within each redundant group. Fig.~\ref{fig:eorpsdiff} shows the fractional difference between the true (input) EoR power spectrum and the estimated power spectrum after gain calibration for a particular 14.6m redundant baseline group. Noticeably, the deviations from the true (input) EoR power spectrum are seen to be almost constant in delay, varying significantly only with LST. This matches the mostly frequency-independent structure of the gain solutions, as shown in Fig.~\ref{fig:gaindiff}.

The typical deviation from the input EoR spectrum is a couple of percent in the most extreme cases, with a few (previously identified) outliers noticeable in isolated LST bins around 15 h. The modulation of the EoR power spectrum can be both positive and negative, so the effect of these calibration errors is different from signal loss/decoherence. The effect is small enough in this case that it is unlikely to have a significant impact on current or near-future upper limits on the EoR signal, although the modulation is likely to induce a spurious component in bispectrum measurements \citep[c.f.][]{Watkinson:2020zqg}. Importantly, because the effect does not depend strongly on delay, there is no unaffected window in delay space where the power spectrum can be extracted without being affected by this modulation.

We observe that the modulation of the EoR power spectrum is dependent on baseline group in essentially the same way as the gain errors, with groups that suffer larger typical gain errors resulting in stronger modulations of the EoR. The gain power spectrum at the beam crossing timescale (see Fig.~\ref{fig:gainps} for the point source + diffuse case) appears to be a reasonably good predictor of the size of the modulation; for example, the Sidelobe 20\% case has the highest gain power spectrum on this timescale, and also one of the strongest modulations in Fig.~\ref{fig:eorpsdiff}. Finally, we note that the difference between the coherently- and incoherently-averaged EoR power spectra is small, and that the isolated `spike' structures in Fig.~\ref{fig:eorpsdiff} are caused by near-zero crossings in the true (input) power spectrum, and so can largely be ignored.

\subsection{Impact of array size}
\label{sec:bigarray}

{
So far in this paper, we have used a relatively small array configuration with 10 antennas. This has limited the maximum baseline length and the number of baselines per redundant group, and has also given rise to some outlier calibration solutions that we would expect to be less common when running the redundant calibration algorithm on larger array configurations. Longer baselines are more chromatic and, according to \citet{Orosz:2018avj}, it is the longer baselines that lead to stronger non-redundant effects on the gain solutions if they are included in the redundant calibration fitting procedure. In this section, we examine some of these issues for two example simulations of a much larger array with 309 antennas.

Our original motivation for restricting our study to a small array was twofold. First, \citet{Orosz:2018avj} found that limiting the redundant calibration procedure to only short baselines significantly reduced the impact of different kinds of non-redundancy. In the preceding analysis we have effectively incorporated this finding into our calibration strategy by default, as only short baselines are available. Second, we have used more complex models for the beams and the sky for our simulations, and a relatively large number of frequency channels and time samples, making the simulation of much larger arrays computationally challenging.

Our large array simulations use the same number of times and frequency channels as before, but now only include the brightest 510 point sources and no diffuse emission. The array specification uses the same dish size and minimum baseline length, but now with 309 dishes arranged into a close-packaged hexagonal pattern. We have considered two different cases: a perfectly redundant array with the fiducial Chebyshev polynomial fit beam (`No perturbation'), and a non-redundant array corresponding to the Stretch Gaussian 2\% case. We have not included noise in these simulations, in order to allow a direct comparison of the baseline-dependent spectral structure in each case. We have also included all baselines in the redundant calibration procedure, including the longest, most chromatic ones.

\begin{figure*}
\begin{center}
\includegraphics[width=0.63\columnwidth]{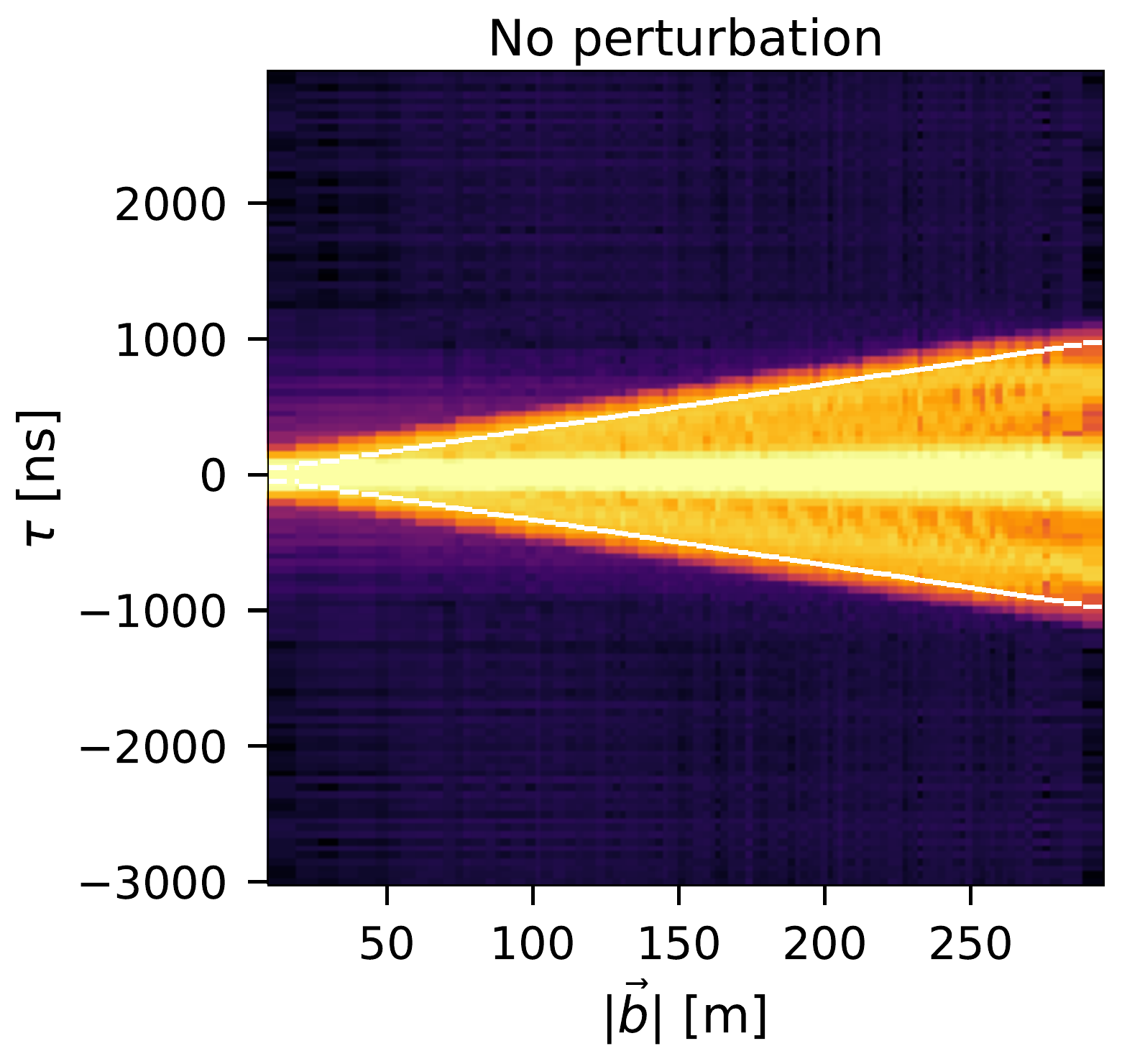}
\includegraphics[width=0.60\columnwidth]{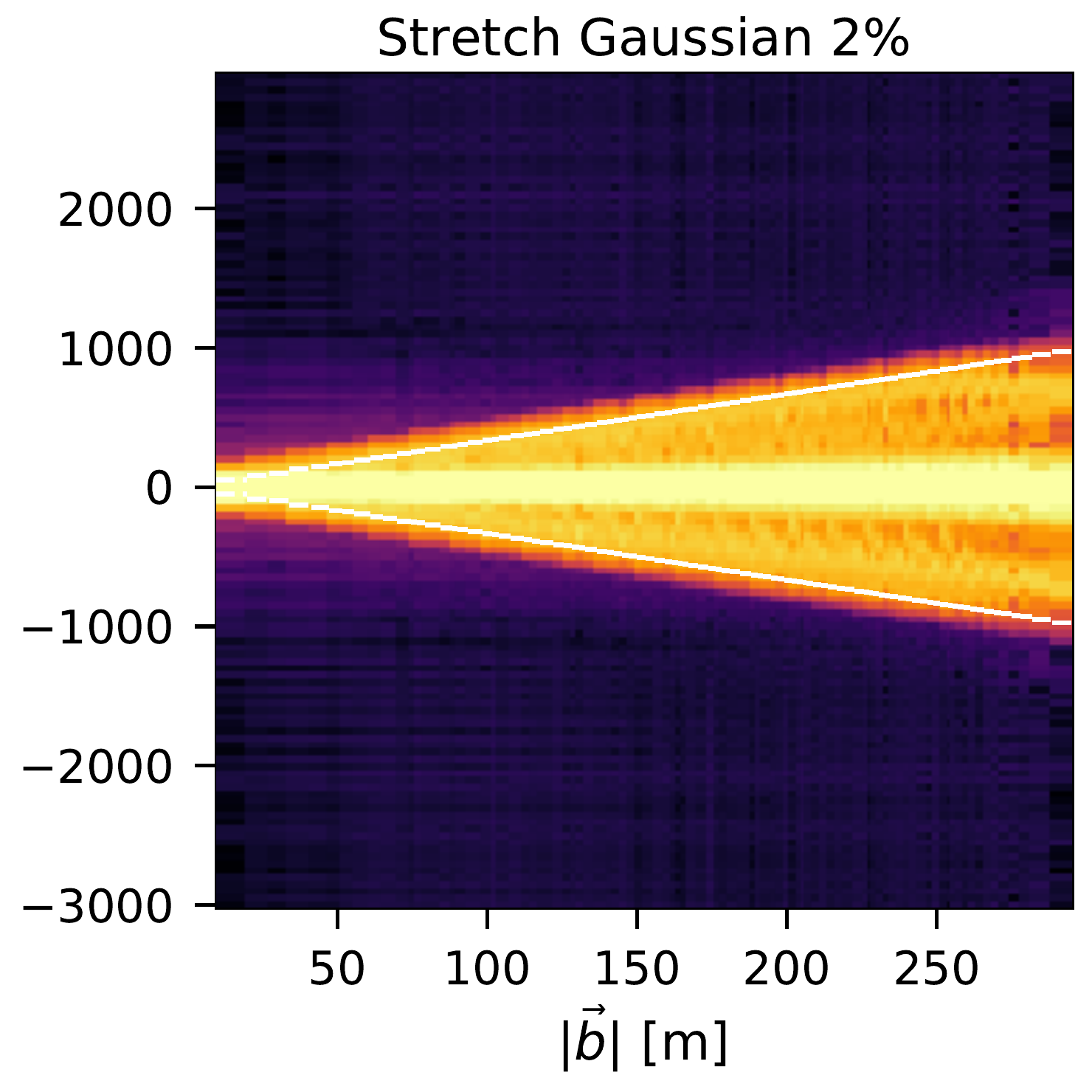}
\includegraphics[width=0.71\columnwidth]{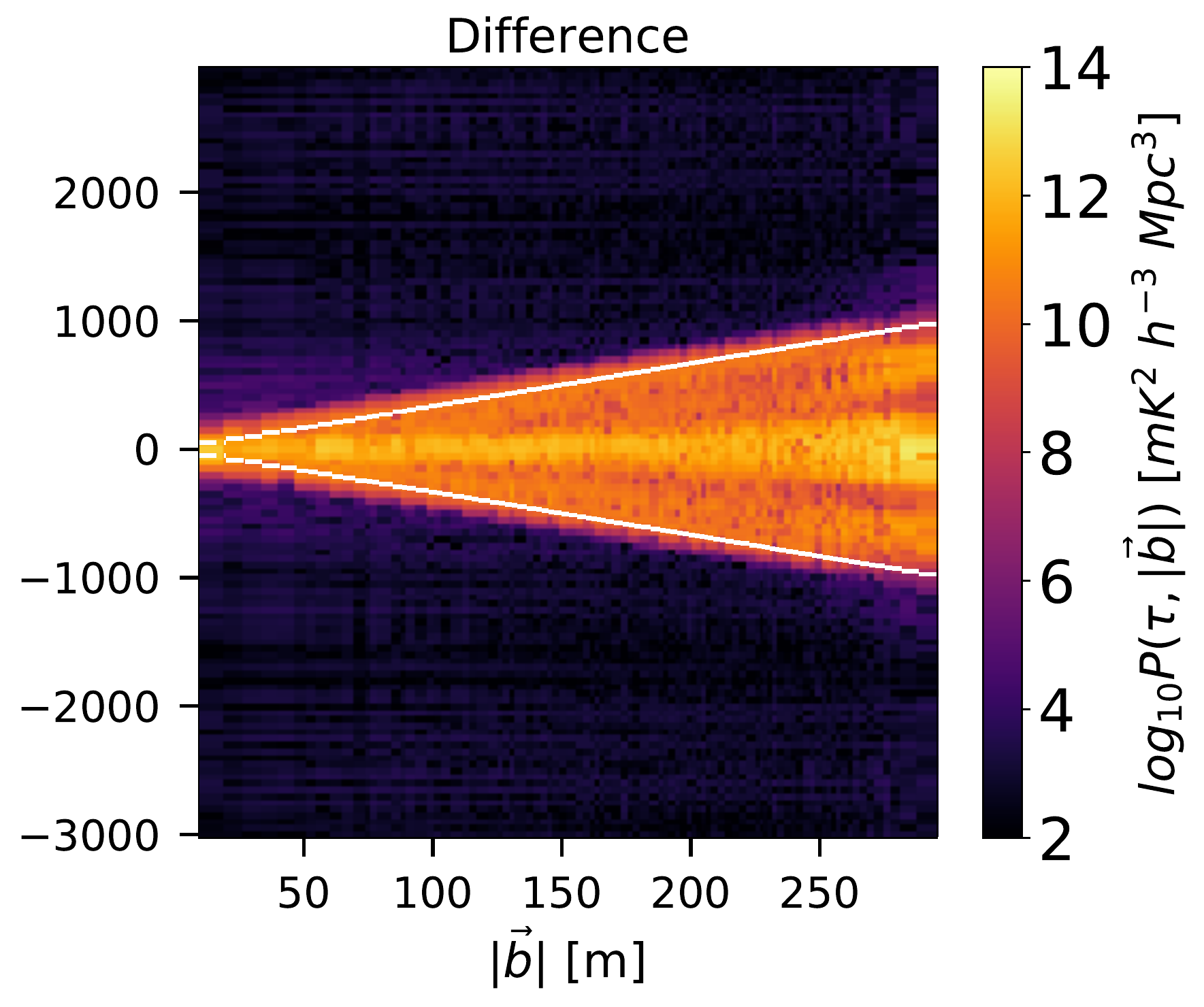}
\caption{{The 2D delay spectrum of the visibilities coherently averaged within each redundant baseline group for a perfectly redundant array with Chebyshev beams (left) and a non-redundant array corresponding to the Stretch Gaussian 2\% case (middle). The absolute value of the difference between the two is shown on the right. All plots are shown for the 309-dish array configuration with no noise, a greatly reduced number of point sources, with no diffuse emission. An incoherent average over all LSTs has also been performed. The white lines shown the horizon limit, which approximately defines the edge of the foreground wedge.}}
\label{fig:wedges}
\end{center}
\end{figure*}

Fig.~\ref{fig:wedges} shows a comparison of the 2D delay spectra for the two cases, following coherent averaging of the visibilities across redundant baseline groups, and incoherent averaging over LST. We have plotted the delay spectra in cosmological units, using the same beam model to calculate the normalisation factor in both cases.

First, we note the structure of the foreground wedge in the `no perturbation' case (left panel). As anticipated, there is a `pitchfork' structure that is characteristic of receivers with dish reflectors, with the worst foreground contamination localised to three forks at $\tau \approx 0$ and the positive and negative horizon lines. There is some leakage outside the horizon into the so-called `buffer' region due to the intrinsic spectral structure of the foregrounds, which is more extensive for the shortest baselines, but there is otherwise little notable structure outside the wedge region, where the amplitude of the power spectrum is very low.

The middle panel of Fig.~\ref{fig:wedges} shows the same quantity, but now for the Stretch Gaussian 2\% case. The resulting 2D power spectrum is quite similar, but differs in small details, as shown more clearly by the difference plot on the right of Fig.~\ref{fig:wedges}. There is a small difference between the two cases across the wedge and buffer regions, owing to the slightly different spectral structure and beam solid angles in the two cases. Most relevant for our purposes however is the small amount of excess power outside the wedge and buffer region, which extends furthest at short and long baseline lengths. This is qualitatively similar to the features seen by \citet{Orosz:2018avj} in their most analogous case ($0.1^\circ$ beam size error, shown in their Fig.~7), although in our case the leakage appears more suppressed at intermediate baseline lengths. The amplitude of the leakage is comparable to the fiducial EoR power spectrum used by \citet{Orosz:2018avj}, although note the higher effective redshift of our simulations ($z \approx 12$, vs. $z \approx 8.5$). As discussed in \citet{Orosz:2018avj}, this leakage can be effectively suppressed by excluding longer baselines from the redundant calibration procedure, which is effectively achieved by default in the smaller (10-antenna) array simulations discussed in previous sections. This leaves the major effect of primary beam non-redundancy on the recovered EoR power spectrum seen in our study as the few-percent modulation effect discussed in Sect.~\ref{sec:eor}.

Finally, we also note the low-level cross-hatched pattern visible outside the wedge region in the three panels of Fig.~\ref{fig:wedges}. We anticipate that this is either numerical noise or low-level ringing caused by a discontinuity in our beam or sky model. The effect is small, below the level of any features that we have studied in this paper, and certainly below the noise level in the 10-antenna simulations. Nevertheless, it can be considered a limitation of these simulations.
}

\section{Discussion and Conclusions}
\label{discuss}

Precise calibration of radio interferometric data is required to detect the 21cm signal from the Epoch of Reionisation in the presence of strong foreground contamination. Redundant calibration is a powerful method for solving for the antenna gains and true (calibrated) visibilities by measuring the same sky mode multiple times, with many baselines of almost-identical length and orientation. Various real-world imperfections in the array, such as positional inaccuracies of the antennas and non-identical primary beams, result in deviations from perfect redundancy. These errors propagate into the redundant calibration solutions, causing gain errors that eventually affect the estimated 21cm power spectrum.

In this paper, we used a suite of visibility simulations to study various types of non-redundancy by perturbing the primary beam of each antenna in a small, close-packaged, hexagonal array with similar properties to the HERA telescope. We considered variations in the mainlobes, sidelobes, and overall stretch, ellipticity, and rotation of each antenna beam, as summarised in Table~\ref{table1}, all based on an analytic fitting model to realistic EM simulations of the HERA primary beam. We compared several different sky models, including a modified GLEAM point source catalogue that covers the whole sky (with and without the brightest sources), and a diffuse foreground model using GSM. To make the simulation more realistic, we also included simple gain fluctuations and an instrumental noise model based on the true sky. With simulations in hand, we then applied a mildly idealised calibration and analysis pipeline to the simulated data to perform a redundant calibration, remove outliers and fix degeneracies in the gain solutions, and finally estimate the delay-space power spectrum.

We presented several different diagnostic quantities in the presence of primary beam non-redundancy, each concerning data products at a different stage of the analysis pipeline. In Sect.~\ref{sec:redgrpvariance}, we measured the variance of the calibrated visibilities within each redundant baseline group following redundant calibration. We saw that the variance increases significantly compared with the perfectly redundant case when the brightest sources are within the field of view, especially if they transit through the mainlobe. For our simulations, the dominant bright source was Centaurus~A, which passes through the mainlobe at LSTs of around $13-14$ h. We saw that the variance was significantly enhanced for the true visibilities (where no gain factors and calibration solutions had been applied) during the transit of Cen~A, but that much of this variability was absorbed by the gain solutions to give the illusion of greater redundancy. For the calibrated data, the intra-redundant group variance was nevertheless significantly higher within this LST range for all types of non-redundancy except for the sidelobe-only perturbation case, when a variance increase was observed between $11-12$ h and $15-16$ h instead (when Cen~A was passing through the sidelobes). There was a broad increase in the variance over the whole LST range when diffuse emission was included in the simulation, but this did not lead to a notable increase in LST-dependent structure; the bright point sources remained the principal cause of such structure.

In Sect.~\ref{sec:gains}, we studied the temporal and spectral structure of the gain errors induced by non-redundancy. The calibration solutions become correlated in time, partially following structure on the sky. This additional correlation appears as a characteristic excess in the temporal power spectrum of the gain solutions on timescales of approximately 20 to 100 minutes, which is comparable to the beam crossing time. This feature in the gain power spectrum is quite similar for all types of non-redundancy, while the gain errors themselves exhibit more characteristic structure between different cases. The errors associated with the transit of Cen~A showed a variety of different behaviours in different cases, for example.

In Sect.~\ref{sec:decorr} we studied a decoherence effect in the power spectrum of coherently-averaged visibilities within a particular redundant group. We defined a decoherence parameter, $\Delta \chi$ (Eq.~\ref{eq:decorr}), that calculates the fractional difference between coherently- and incoherently-averaged power spectra as a function of LST, with reference to the time-averaged incoherently averaged power spectrum. We found that $\Delta \chi$ deviates most significantly from zero when the brightest source, Cen~A, is passing close to the mainlobe (at LSTs of around $13-14$ h), and for some types of non-redundancy can be as large as $-2.5\%$. In comparison, the maximum value was $-14\%$ when the true (simulated) visibilities were considered, which suggests that the redundant calibration reduces the decoherence effect due to the intrinsic non-redundancy of the data in some circumstances.

Other bright sources appear to have comparatively little effect on this statistic, although we do see an increase in LST-dependent fluctuations in $\Delta \chi$ when diffuse emission is included in the simulations. There was also a significant baseline orientation effect in the observed decoherence, with differences in the timing of peak decoherence between baselines of the same length but different orientations. This suggests a possible route towards identifying different types of primary beam non-redundancy in real data, based on how the decoherence statistic evolves with LST as a very bright source transits. Longer baselines were also more strongly affected by decoherence, although the small size of our simulated array prevented us from studying more than a handful of different baseline lengths.

Finally, in Sect.~\ref{sec:eor} we applied the redundantly-calibrated gain solutions to a simulated EoR component, finding that the recovered EoR power spectrum was modulated as a function of LST, but that no significant additional structure was induced as a function of delay. {This conclusion is based on the delay spectrum approach given in \citet{Parsons:2012qh}. However, the effect of the calibration error might change depending on the choices of other power spectrum estimator \citep{morales19}.} While this effects is relatively small in the cases we studied, and so is not likely to significantly affect current and near-future upper limits on the EoR power spectrum, we do expect it to contaminate measurements of the signal bispectrum.

Taking all of these diagnostic quantities together, we found that the type of non-redundancy where there is a single outlier antenna with a significantly different primary beam sustains the most severe gain errors. {A similar study of broken MWA dipole antennas showed that this introduces a bias in the cylindrical power spectrum of the order of $\sim10^3 {\rm mK}^2 {\rm h}^{-3} {\rm Mpc}^3$ \citep{joseph20}.} This suggests the possibility of making relatively easy improvements in redundant calibration quality by identifying and excluding the antennas with the most discrepant primary beams.  Non-redundancy in the sidelobes did not generally cause severe gain errors, although it did result in stronger modulations of the recovered EoR spectrum. Differences in ellipticity between antennas caused substantial gain errors, although this depends on the distribution of ellipticity and rotation; relatively small rotations of the beam result in quite minor errors for example.

Overall, the most severe effects were associated with the brightest sources that pass close to the field of view, suggesting that a focus on modelling these sources and the primary beam response to them should be sufficient to disentangle the worst effects of primary beam non-redundancy. We further found that diffuse emission (away from the Galactic plane) does not impart significant additional structure in the gain errors in most cases aside from increasing the overall noise level, although noticeable effects did start to creep in as the Galactic plane began to rise.

In many ways, these results are reassuring -- our findings suggest that relatively modest additional modelling and data cuts should be enough to mitigate the worst effects of non-redundant primary beams. This is only part of the picture for real instruments however, which also suffer from (e.g.) antenna position and polarisation non-redundancies that we did not study here. These can give rise to effects that couple in different ways to the sky, different dependencies on baseline length and orientation, {different frequency dependencies} etc. We also used a reasonably small array with only 10 antennas for reasons of computational efficiency, and so are missing effects that predominantly affect longer baselines, such as those identified by \citet{Orosz:2018avj}.

In another simplification, we did not incorporate a realistic sky-based absolute calibration step in our simulated pipeline, which would have likely substantially altered the characteristics of the gain errors. Finally, we did not include the effects of polarisation leakage, which will also couple the redundant gain solutions to the sky in complex ways.

{To alleviate some of these caveats, we investigated a couple of illustrative examples of larger (309-dish) arrays in Sect.~\ref{sec:bigarray}, and studied an extended primary beam model with frequency-dependent non-redundant sidelobe perturbations in Sect.~\ref{sec:freqsl}. We also examined the effects of temporal gain smoothing in Sect.~\ref{sec:gainsmooth}, which more closely mirrors how redundant calibration is applied to real data when combined with an absolute calibration step \citep[c.f.][]{2020ApJ...890..122K}. For the frequency dependence of the perturbations in particular, we note that there are very many possible combinations of ways to perturb the angle, frequency, and antenna dependence of the primary beams. In order to keep the number of scenarios tractable, future studies will need to use more specific and realistic functional forms for these perturbations than we have considered here.}

\section*{Acknowledgements}

We are grateful to James Aguirre, Josh Dillon, Bryna Hazelton, Nick Kern, Piyanat Kittiwisit, Adrian Liu, Steven Murray, and Saurabh Singh for useful comments and discussions, {and the anonymous referee for useful comments}.

SC is supported by a research contract between Queen Mary University of London and the University of California at Berkeley. PB and HG acknowledge support from STFC grant ST/T000341/1. The software used to generate the results in this paper is available from \url{https://github.com/philbull/non-redundant-pipeline/}.

\section{DATA AVAILABILITY}
The data from this study  will be shared on reasonable request to the corresponding author.

\balance

\bibliographystyle{mnras}
\bibliography{primarybeam} 

\end{document}